\documentclass[11pt]{article}

\usepackage{palatino,setspace,graphicx,rotating,mathrsfs,bm,listings,caption,natbib}
\usepackage{subfigure,latexsym,amsmath,amssymb,natbib,rotating,ocg-p}
\usepackage{algorithm, algorithmicx, xparse}
\usepackage[noend]{algpseudocode}

\newcommand{\XX}{\mbox{$\mathbf X$}}
\newcommand{\YY}{\mbox{$\mathbf Y$}}
\newcommand{\Set}{\mbox{$\mathbf S$}}

\makeatletter
\long\def\@makecaptiohen#1#2{%
   \vskip\abovecaptionskip
   \sbox\@tempboxa{#1: #2}%
   \ifdim \wd\@tempboxa >\hsize
     #1: #2\par
    
   \else
     \global \@minipagefalse
     \hb@xt@\hsize{\box\@tempboxa\hfil}%
   \fi
   \vskip\belowcaptionskip}

\makeatother

\newcommand{\xx}{\mathbf{x}}
\newcommand{\yy}{\mathbf{y}}
\newcommand{\vv}{\mathbf{v}}

\begin{document}


\begin{titlepage}
\begin{singlespace}

\vspace*{.8in}
\begin{center}
\Large {\sc
    \textbf{A Parallel Evolutionary Multiple-Try Metropolis Markov
      Chain Monte Carlo Algorithm for Sampling Spatial Partitions}}
  \vspace{0.8cm}
        
  \normalsize Wendy K. Tam Cho\footnote[1]{ Department of Political
    Science, Department of Statistics, Department of Mathematics,
    Department of Asian American Studies, the College of Law, and the
    National Center for Supercomputing Applications, University
    of Illinois at Urbana-Champaign.  {\em wendycho@illinois.edu}} \\
  \vspace*{.1cm} Yan Y. Liu\footnote[2]{ Research Scientist,
    Computational Urban Sciences Group, Computational Science and
    Engineering Division, Oak Ridge National Laboratory.  {\em
      yanliu@ornl.gov}} 
  \renewcommand{\thefootnote}{\arabic{footnote}}

\end{center}

\vspace{2mm}
\begin{abstract}
\noindent
We develop an Evolutionary Markov Chain Monte Carlo (EMCMC) algorithm
for sampling spatial partitions that lie within a large and complex
spatial state space.  Our algorithm combines the advantages of
evolutionary algorithms (EAs) as optimization heuristics for state
space traversal and the theoretical convergence properties of Markov
Chain Monte Carlo algorithms for sampling from unknown distributions.
Local optimality information that is identified via a directed search
by our optimization heuristic is used to adaptively update a Markov
chain in a promising direction within the framework of a Multiple-Try
Metropolis Markov Chain model that incorporates a generalized
Metropolis-Hasting ratio.  We further expand the reach of our EMCMC
algorithm by harnessing the computational power afforded by massively
parallel architecture through the integration of a parallel EA
framework that guides Markov chains running in parallel.
\end{abstract}

\vspace{35mm}
\end{singlespace}

\end{titlepage}

\pagestyle{headings}

\section{Introduction}

Markov Chain Monte Carlo (MCMC) methods originated in statistical
physics~\citep{Metropolisetal:53} and have migrated to applications in
many disciplines.  A particular use of MCMC that has wide application
is for sampling from complex and unknown distributions.  While the
theory of MCMC methods ensures sampling from unknown distributions, it
is not always straightforward to devise these methods for a particular
application, and the theoretical result, moreover, is asymptotic.
Indeed, if the application problem is large and the state space is
difficult to traverse, the amount of time required before the
theoretical convergence of a Markov chain is realized may be
prohibitively long.  Hence, while MCMC methods are theoretically
attractive, successful implementation for complex applications can be
quite challenging.

\begin{figure}
\centering
\includegraphics[width=3in]{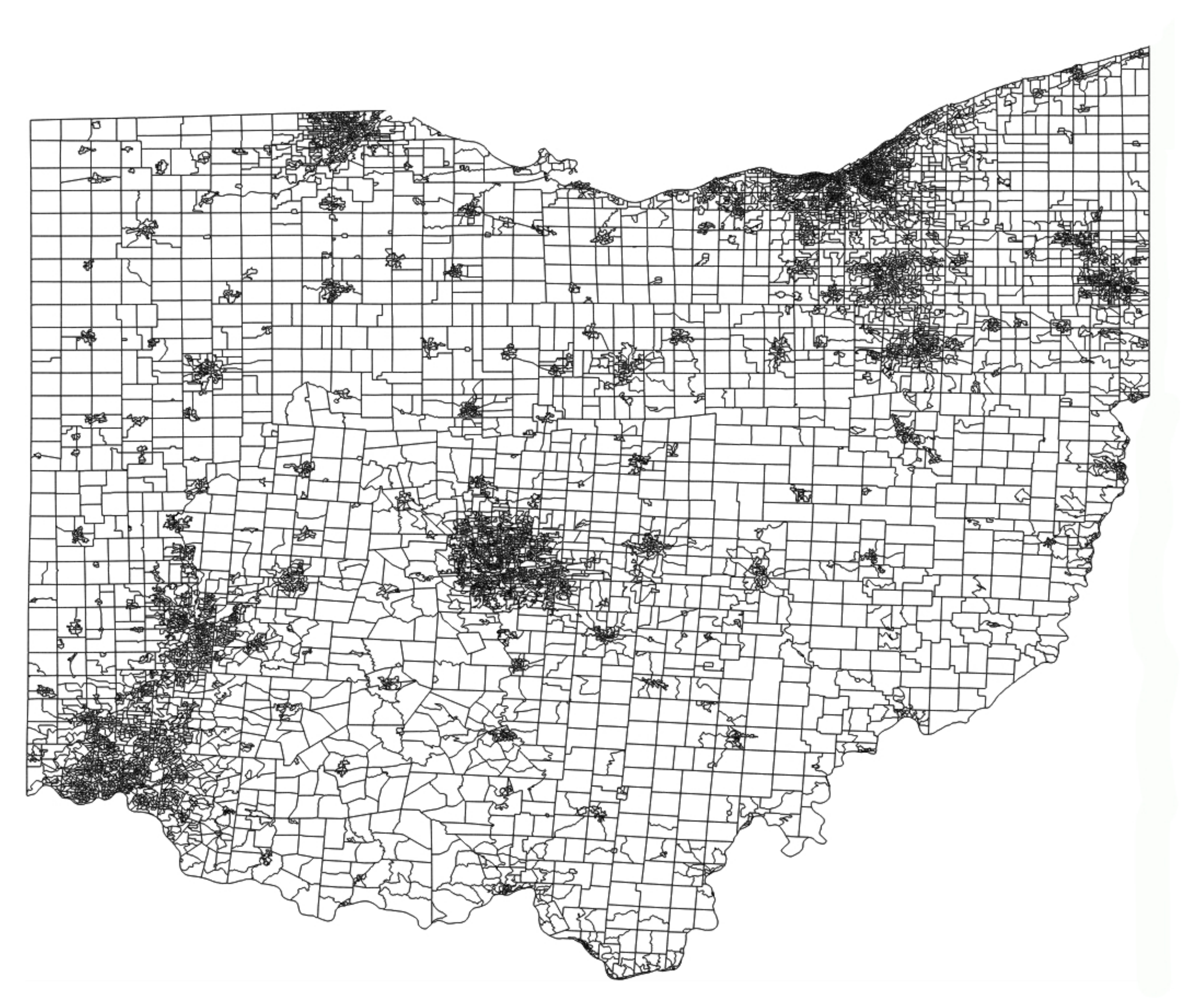}
\includegraphics[width=3in]{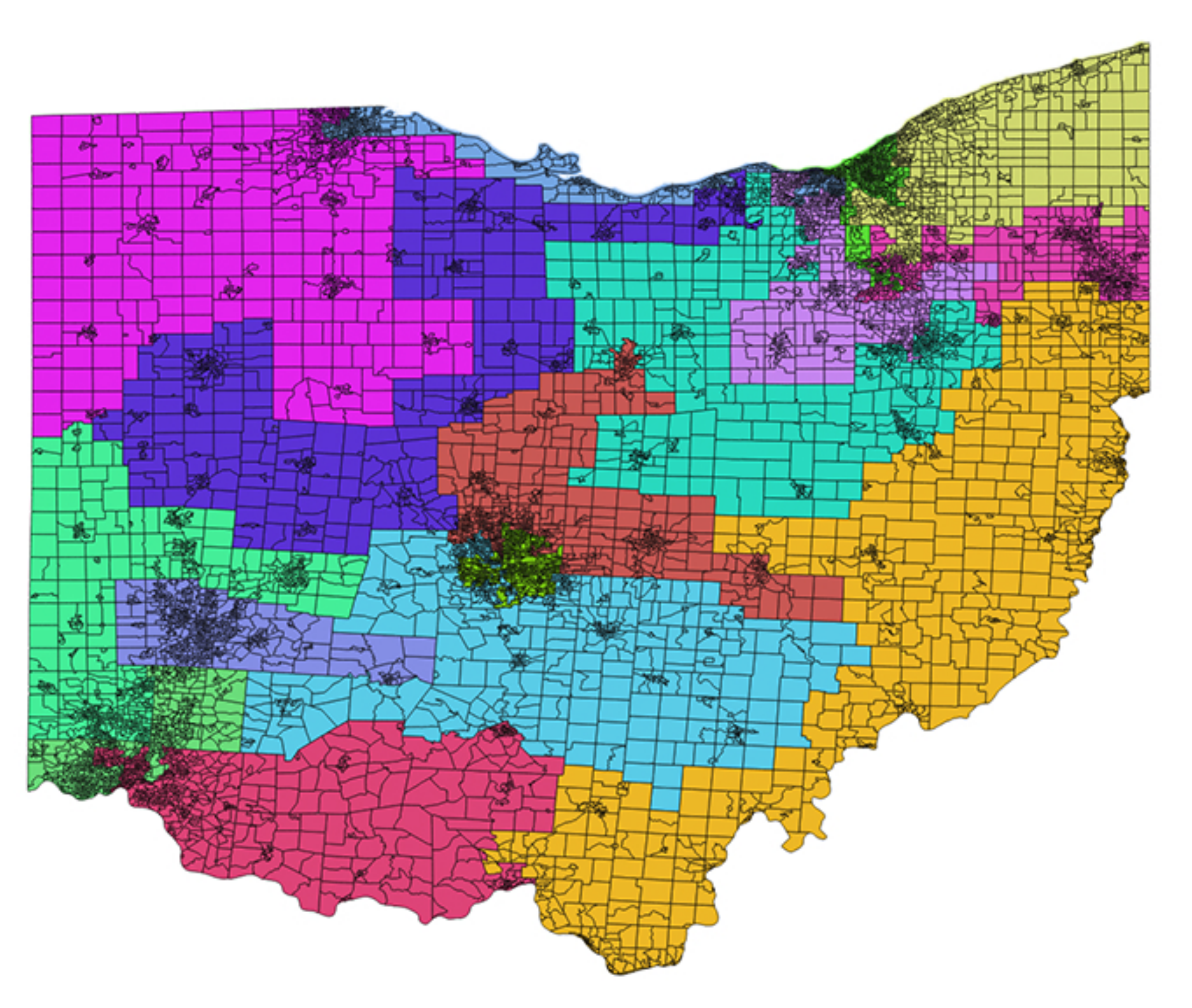}
\caption{Spatial Geographic Units that Encompass the State of Ohio and
  a Particular Spatial Partition of those Units into 16 Disjoint
  Contiguous Zones}
\label{fig:ohio}
\end{figure}

Consider the spatial partitioning application illustrated in
Figure~\ref{fig:ohio}.  The goal is to characterize the set of all
possible feasible partitions of size $k$, where feasible is defined as
a partition that satisfies a specific set of constraints (e.g.,
contiguity).  The plot on the left displays a fixed set of geographic
units that collectively encompasses the entire state of Ohio.  The
plot on the right shows one particular mutually exclusive and
exhaustive partition of these geographic units into $k=16$ disjoint
contiguous sets.  One can visually see that, given the large number of
base geographic units, there is an enormous number of distinct
partitions into 16 disjoint contiguous sets.  Indeed, this number is
sufficiently large that it is computationally infeasible to enumerate
the set of all partitions within current computing capabilities.
Since an exhaustive search is not possible, an alternative route is to
produce a representative sample of the universal set, which would also
allow us to derive insights into the underlying population of feasible
partitions.  MCMC methods provide a possible route for sampling from
this unknown population distribution.  However, how to specify an MCMC
model that produces a sample of feasible spatial partitions is not
obvious.

For many applications, a common MCMC strategy is to define a Markov
transition function that amounts to a small or local random change in
the current state.  Small or local changes are attractive for two
reasons.  First, they are conceptually and operationally
simple---these moves are easy to define and relatively simple to
implement.  Second, the Metropolis-Hastings
ratio~\citep{Metropolisetal:53, Hastings:70} needs to regularly lead
to accepted transition proposals.  Since a small change likely results
in a large Metropolis-Hastings ratio, the movement of the Markov chain
is then fairly fluid.  At the same time, because these are small
movements in a very large state space, the resulting Markov chain
converges slowly and is, moreover, likely to become trapped in
localized regions.  Hence, for large or complex applications, though
this type of chain may be simple to specify, it is not likely to
converge rapidly enough to be practically useful.  In this case,
performance is compromised for conceptual ease.

For the spatial partitioning application, given a feasible state, a
small or local random change might be to choose a geographic unit on a
zone border and reassign it to an adjacent zone.  This type of
movement is conceptually and operationally simple, both to understand
and to implement.  One can see, however, that when there are a large
number of geographic units, relying solely on this type of movement to
traverse the underlying population of feasible partitions would be
inefficient, and likely prohibitively so.

To improve performance and hasten convergence, one approach is to
define larger steps for the Markov chain.  Two particular benefits
accrue from pursuing this strategy.  First, larger moves may enable a
chain to escape from local optimal states.  Second, if well-designed,
these larger moves can more efficiently and effectively traverse a
large state space, leading to faster convergence of the chain.
However, how to conceive intelligent and large movements that
effectively transition from one feasible partition to another feasible
partition is not easily discernible.  Simply ``large'' movements are
likely to result in small Metropolis-Hastings ratios, which lead to
rejected proposals, and thus to a non-fluid and ineffective Markov
chain.

Plainly, a central goal for an effective MCMC algorithm is to devise a
chain that is able to traverse the state space in both an effective
and efficient manner.  Notably, effective and efficient solution space
traversal is precisely the same objective that animates the research
on optimization heuristics.  Here, we examine how insights that have
been gleaned from the optimization heuristics literature might be
adapted to improve the performance of MCMC algorithms for the spatial
partitioning problem.  The main task in marrying these two literatures
is to fit the mechanics of the optimization search within the MCMC
theoretical framework that enables sampling.

\section{Intelligent State Space Traversal}

\subsection{Gibbs Sampler}

Intelligent and randomized state space traversal are fundamental to
performance, whether for an optimization heuristic or for an MCMC
algorithm.  One way in which the moves can be both random and
purposeful rather than simply random, is to follow the local dynamics
of the target distribution.  A Gibbs sampler follows the local
dynamics of the target distribution by composing a sequence of
conditional distributions along a set of sampling directions.  In
practice, while Gibbs samplers are ensured theoretical convergence,
they may converge quite slowly, because it is often not clear how to
intelligently sample from the full conditional distributions.

In the Gibbs sampler, we wish to sample from $f(\mathbf{x})$ where
$\mathbf{x}$ can be broken down into components,
$\mathbf{x} = (x_1, x_2, \dots, x_n)$.  One iteration of a systematic
scan Gibbs sampler can be described as follows.  Begin at some state,
$\mathbf{x}^{(0)}$.  At iteration $i$,
\begin{enumerate}
\item[1.] Sample $x_1^{(i+1)}$ from the conditional distribution
  $f(x_1 \, | \, x_2^{(i)}, x_3^{(i)}, \dots, x_n^{(i)})$.
\item[2.] Sample $x_2^{(i+1)}$ from the conditional distribution
  $f(x_2 \, | \, x_1^{(i+1)},
  x_3^{(i)}, \dots, x_n^{(i)})$. \\
  $\vdots$
\item[n.] Sample $x_n^{(i+1)}$ is sampled from
  $f(x_n \, | \, x_1^{(i+1)}, x_2^{(i+1)}, \dots, x_{n-1}^{(i+1)})$
\end{enumerate}
The sequence of realizations,
\{$\xx^{(t)} = (x_1^{(t)}, x_2^{(t)}, \dots, x_n^{(t)}) \}$, forms a
Markov chain with stationary distribution, $f(\xx)$.

Two challenges must be overcome for the Gibbs sampler to be
successful.  First is the issue of {\em conjugacy} or the ability to
sample from the conditional distributions~\citep{CarlinGelfand:91}.
Second is how to overcome the difficulty in constructing univariate
sampling directions in such a way that ensures rapid movement around
the support of $f(\xx)$ when simple random movement is insufficient.
In other words, while it is possible to devise a Gibbs sampler that
follows the local dynamics of the target distribution, how precisely
one intelligently, and thus effectively, devises this movement is
non-obvious and, further, application dependent.

The conjugacy issue can be addressed with the Griddy-Gibbs
sampler~\citep{RitterTanner:92} where rather than sampling from the
full conditional distributions, one may form a simple approximation of
the inverse CDF on a grid of points.  However, even with a successful
implementation of the Griddy-Gibbs sampler that addresses the
difficulties of sampling from the full conditional distributions, the
challenge of devising intelligent movements in purposeful directions
remains.

\subsection{Adaptive Direction Sampling}

Adaptive Direction Sampling (ADS) is a general technique that fits
within the framework of the Gibbs sampler and aims to sample in a
particular search direction.  It was introduced as an automated way of
overcoming the slow convergence in the Gibbs sampler by providing a
way to guide sampling from the conditional
distributions~\citep{Gilksetal:94, RobertsGilks:94}.  The ADS
algorithm also introduces a way in which the interaction of multiple
Markov chains can be specified to improve the performance of MCMC by
proposing that the sampling direction be updated with information from
a different Markov chain state.

ADS can be adapted in various ways.  The general form of ADS is
\begin{equation}
\xx_c^{(t+1)} = \xx_c^{(t)} + r (\vv^{(t)} + u^{(t)} \xx_c^{(t)}) ,
\label{eq:ads}
\end{equation}
where $\vv^{(t)}$, an $n$-vector, and $u^{(t)}$, a scalar, are any
functions of the current set, $\Set(t)$, excluding $\xx_c^{(t)}$, the
current state, all at time or state $t$.  If $\vv^{(t)}$ is the
difference between two points in the current set, parallel ADS
emerges.  If $\vv^{(t)}$ is a random coordinate direction with
$u^{(t)} = 0$, we have the Gibbs sampler.  The hit-and-run algorithm
emerges when $\vv^{(t)}$ is a random direction and
$u^{(t)} = 0$~\citep{Belisleetal:93}.

Although adapting the direction of the sampling with information from
another state can clearly be helpful, the core problem with
intelligent state space traversal remains.  Indeed, studies of the ADS
algorithm indicate that it is not particularly effective for improving
sampling efficiency generally if the direction generated by ADS
remains rather arbitrary.  Instead, moving in the direction of the
other chains improves the state space traversal only when such
movement coincides with transitions that are intelligent and
purposeful.

\subsection{Adaptive Direction Sampling Guided by Evolutionary
  Algorithm Operators}

Rather than relying on an arbitrary direction, \citet{LiangWong:01}
guide the directional sampling of ADS with their {\em snooker
  crossover}, a recombination operator from an evolutionary algorithm
(EA) that they modify to fit within the MCMC framework.  That is, they
utilize an optimization heuristic to define the sampling direction in
ADS.  In the snooker algorithm, a set of possible states, $\Set$, is
retained.  In the language of evolutionary algorithms, these possible
states comprise the population.  From the set, $\Set$, two states are
chosen randomly, one as the current state, $\xx_c$, and one as the
anchor state, $\xx_a$.  The direction of movement is then chosen along
a line that connects the current state and the anchor state.  From
equation (\ref{eq:ads}), the snooker algorithm arises when
$u^{(t)} = -1$ and $\vv^{(t)} = \xx_a^{(t)}$, where $\xx_a^{(t)}$ is a
randomly chosen state from $\Set(t)$, excluding $\xx_c^{(t)}$.  To
prevent the new states from being too concentrated around the anchor
state, one can sample from an adjusted full conditional distribution.
The intuition is that after a burn-in period, the set of states is
likely to be in high density regions, and high density states are
helpful for identifying sampling directions for a Gibbs sampler.

While the snooker algorithm is a recombination evolutionary algorithm
operator that can be adapted to fit within the mathematical framework
of MCMC, the effectiveness and efficiency of the algorithm for
recovering samples for any particular application is still dependent
on how its operators traverse that {\em particular} state space.  In
the optimization literature, it is clear that while EAs have been
successful for many applications, significant effort must still be
expended to adapt the heuristic to the particular solution landscape.
To be sure, EAs are not panaceas.  Rather, they comprise a general
technique whose success is dependent on successful adaptation of the
evolutionary framework to a specific application.

All the same time, combining optimization heuristics within an MCMC
framework is a promising strategy for difficult problems, and one that
has already been successfully implemented for a number of interesting
problems, including Bayesian mixture models, $C_p$ model sampling, and
change point problems~\citep{LiangWong:00, LiangWong:01,
  LaskeyMyers:03}.  The main hindrance is that the approach is
general, and successful implementation of both the optimization as
well as the MCMC components, especially for complex applications, must
be tuned to the idiosyncrasies of the application state space.

\subsection{The Multiple-Try Method}
\label{subsec:MTM}

\citet{Liuetal:00} generalize how Markov chain movement can be adapted
with optimization heuristics in their Multiple-Try Method (MTM).  They
propose a way to incorporate optimization steps into an MCMC sampler
with a multiple-try Metropolis-like transition rule that enables a
more thorough exploration of the neighboring region that is defined by
a transition proposal, $T(\xx,\yy)$.  The method is general
and~\citet{Liuetal:00} demonstrate how MTM can be combined with
different MCMC variants (in particular, ADS, Griddy-Gibbs samplers,
and the Hit-and-Run algorithm) to produce more effective samplers.

In the MTM framework, suppose  that the current state is $\xx$.  The
optimization heuristic proposes some set of $m$ proposal moves,
$\YY = \{\mathbf{y}_1, \dots, \mathbf{y}_m \}$.  In this way, the
optimization heuristic provides a method for directional sampling
without the shortcoming of previous approaches that relied primarily
on a random direction.  Importantly, since an optimization move may
not be a proper Markov transition, it cannot simply be accepted as the
next state in a Markov chain.  Instead, from the set of proposal
states created by the optimization heuristic, one proposal state,
$\yy \in \YY$, is selected probabilistically, and this selected
proposal is then accepted or rejected according to a generalized
Metropolis-Hastings ratio that has the form,
\begin{equation}
  r_g = \min \left\{ 1, \frac{\xi (\mathbf{y}_1, \mathbf{x}) + \dots +
      \xi (\mathbf{y}_m,\mathbf{x})}{\xi (\mathbf{x}_1^*,\mathbf{y}) +
      \dots + \xi (\mathbf{x}_m^*, \mathbf{y})} \right\} ,
\end{equation}
where $\xi (\mathbf{y}, \mathbf{x}$) is the probability of choosing
proposal $\mathbf{y}$ from the set $\YY$, and
$\{ \xx_i^*\}, i = 1, \dots, m$, comprise a set of proposal moves from
the same optimization heuristic that now begins at state $\yy$ and
moves in the direction of state $\xx$.

\citet{Liuetal:00} develop and demonstrate these ideas with a
Conjugate-Gradient Monte Carlo where the search direction proceeds
along a vector whose direction is based on the conjugate gradient.
Once the direction is chosen, randomness is injected via the magnitude
of this vector.  Searching along a gradient is one optimization
technique, but plainly not the only option.  This tactic is effective
for certain types of state spaces, but not ideal for other types, and
certainly not for all types.  Indeed, optimization heuristics come in
many different genres, and within these genres, with many different
specifications and implementations.  However, the MTM framework is
general such that {\em any} optimization technique can be employed to
create a set of proposal moves that are then accepted or rejected
according to the generalized Metropolis-Hastings ratio.

The evolution of these algorithms highlights a theme encountered
repeatedly both in statistical modeling and in the design of
optimization heuristics.  Namely, there is not one fixed solution that
is ideal for every application.  There are only general guiding
principles for utilizing particular modeling frameworks.
Incontrovertibly, devising Markov chains must be done with the
peculiarities of the application in mind.  As well, the success of
optimization heuristics increases in likelihood when they are designed
to adapt to application idiosyncrasies.  In every case, circumspection
is necessary in the selection of a meaningful and effective search
direction for either the Markov chain transitions or the optimization
heuristic steps.

\section{An Evolutionary Algorithm for Spatial Optimization}

While traversal of some state spaces may seem to fit into standard
frameworks, other state spaces are more unusual.  Constrained spatial
state spaces produce particularly thorny applications since it is
difficult to envision how one might transform multi-dimensional
spatial characteristics into uni-dimensional searches.  Indeed,
spatial constraints pose particular problems for both spatial
optimization as well as sampling techniques because spatial
requirements can impose significant restrictions on the viability of
large portions of the possible solution space.~\citep{Duqueetal:11,
  FrancisWhite:74, HofBevers:98, OKellyMiller:94}

In the context of evolutionary algorithms, we know that conventional
EA heuristics, based on non-spatial binary problem encodings, prove to
be problematic for spatial optimization problems.  Indeed, for spatial
applications, the efficiency and effectiveness of an algorithm is
closely tied to how the spatial properties are integrated into the
decision search space~\citep{LiuCho:20}.  In particular, when spatial
characteristics are not explicitly considered in the algorithm design,
the resulting EA is inefficient and expends significant computational
effort exploring infeasible portions of the search space where the
spatial constraints are violated.  The severity of this problem
increases quickly with problem size.

The spatial partitioning optimization problem is closely related to
our spatial sampling problem.  In previous work, we have presented an
EA for addressing the spatial and computational complexities of this
related optimization problem~\citep{Liuetal:16, ChoLiuSC:17,
  LiuCho:20}.  We now review this optimization heuristic and delineate
how it can be integrated into an MCMC framework that enables the
ability to sample spatial partitions.

\subsection{Spatial Partitioning Application}

Consider the spatial partitioning application where we wish to
partition a set of $n$ spatial units, $u_1, u_2, \dots, u_n$, into a
set of $k$ disjoint spatially contiguous zones that each
satisfy a specified set of spatial as well as non-spatial constraints.

\vspace*{3mm}
\noindent 
{\bf Mathematical Formulation}
\vspace*{-8mm}

\begin{singlespace}
\begin{quote}
 $I$: set of spatial units;
 
 $A$: set of adjacent unit pairs;
 
 $K$: set of zones;
 
 $n_k$: number of units in zone $k$;

$\xx = \lbrace x_{ik} \rbrace: x_{ik} = $
\begin{math}
  \left\{
    \begin{array}{l}
      1 \hspace{2mm} \mbox{if unit} \; i \; \mbox{is assigned to zone} \; k \\
      0 \hspace{2mm} \mbox{otherwise}
    \end{array}
  \right.
\end{math}
 
 $y_{ijk}$: flow from unit $i$ to unit $j$ for zone $k$
 
 $h_{ik} = $
\begin{math}
  \left\{
    \begin{array}{l}
      1 \hspace{2mm} \mbox{if unit} \; i \; \mbox{is the hub of
      zone} \; k \\
      0 \hspace{2mm} \mbox{otherwise}
    \end{array}
  \right.
\end{math}
\end{quote}

Objective  \hspace*{2.2cm} minimize $\mbox{{\sl obj}}(\xx)$

Constraints 
\begin{eqnarray}
\displaystyle\sum_{j \, \mid \, (i,j) \, \in A}{y_{ijk}} - \sum_{j \, | \, (j,i) \ \in
  A}{y_{jik}} = n_{k}h_{ik} - x_{ik} \hspace*{36mm} \forall \, k \in K, \; \forall \, i \in I
  & &  \label{constraint1} \\
\displaystyle\sum_{j \, | \, (j,i) \, \in A}{y_{jik}} \leqslant (n_{k} - 1)
  x_{ik} \hspace*{62mm} \forall \, k \in K, \; \forall \, i \in I   & & \label{constraint2} \\
\displaystyle\sum_{k \in K} x_{ik} = 1 \hspace*{99mm} \forall \, i \in I & & \label{constraint3} \\
\displaystyle\sum_{i \in i} h_{ik} = 1 \hspace*{96.5mm} \forall \, k \in K & & \label{constraint4} \\
\displaystyle a \xx \leqslant b \hspace*{114mm} \mbox{}& & \label{constraint5} \\
x_{ik}, h_{ik} \in \lbrace 0,1 \rbrace \hspace*{73mm} \forall \, k \in K, \forall \, i
  \in I & &  \label{constraint6} \\
y_{ijk} \geqslant 0 \hspace*{79mm} \forall \, k \in K, \forall \, (i,j) \in A & & \label{constraint7}
\end{eqnarray}
\end{singlespace}

\vspace*{5mm} 
The above formulation is revised from~\citet{Shirabe:09}, who defines
the contiguity for each partition as a network flow from all of the
spatial units in each zone to their zone hub that receives the flow.
The objective function, {\sl obj}(), is a weighted sum of spatial and
non-spatial objectives. Constraint (\ref{constraint1}) requires that
the difference of flow into a unit $i$ and out of $i$ must be
$(n_k - 1)$.  This means that if unit $i$ is the hub of zone $k$, the
flow traverses each unit in the zone exactly once.  Otherwise, this
constraint has no effect.  Constraint (\ref{constraint2}) ensures that
no unit is visited twice.  Constraint (\ref{constraint3}) guarantees
that each unit is a member of one and only one zone.  Constraint
(\ref{constraint4}) guarantees that each zone has only one hub.  These
four constraints together ensure that all of the units are partitioned
into exactly $k$ contiguous zones.  Constraint (\ref{constraint5})
denotes all other non-spatial constraints.  While unit assignment is
discrete (Constraint (\ref{constraint6})), the flow is formulated as a
continuous variable (Constraint (\ref{constraint7})).  We can also see
that this problem is computationally intractable since Constraints
(\ref{constraint1}) and (\ref{constraint2}) generate a number of
inequalities that increases exponentially with the number of units.

One example of a constraint is the requirement for weight balancing.
across zones. Here, each spatial unit, $u_j, j = 1, \dots, n$, is
associated with a weight, $u_j^{(w)}$.  Each zone, $i = 1, \dots, k$, is
an aggregation of some number of these units, and the weight for zone
$i$ is $w_i = \sum_{u_j \in \, \mbox{\scriptsize{zone}} \, i} u_j^{(w)}$.
The aggregated weight for any one zone is required to be within a
small specified range, $\varepsilon$, of all other zones,
\begin{equation}
\label{eq:weightbal}
\frac{\max_{i=1}^k {w_i} - \min_{i=1}^k {w_i}}{\sum_{i=1}^k w_i} < \varepsilon .
\end{equation}

There are many other constraints that may be imposed on the solutions.
One example is a mathematically defined shape requirements that may,
for instance, constrain the size of the isoperimeter quotient or the
ratio of the area of a zone to the area of a circle with the same
perimeter.  Another example of a spatial constraint is a requirement
that certain spatial units must be contained in the same zone.  While
there are a large number of possibilities for spatial constraints,
they all have the effect that they limit the global state space to a
smaller feasible state space where feasibility is defined as states
that satisfy these specific spatial and non-spatial constraints.

\begin{figure}
\centering
\includegraphics[width=4.5in]{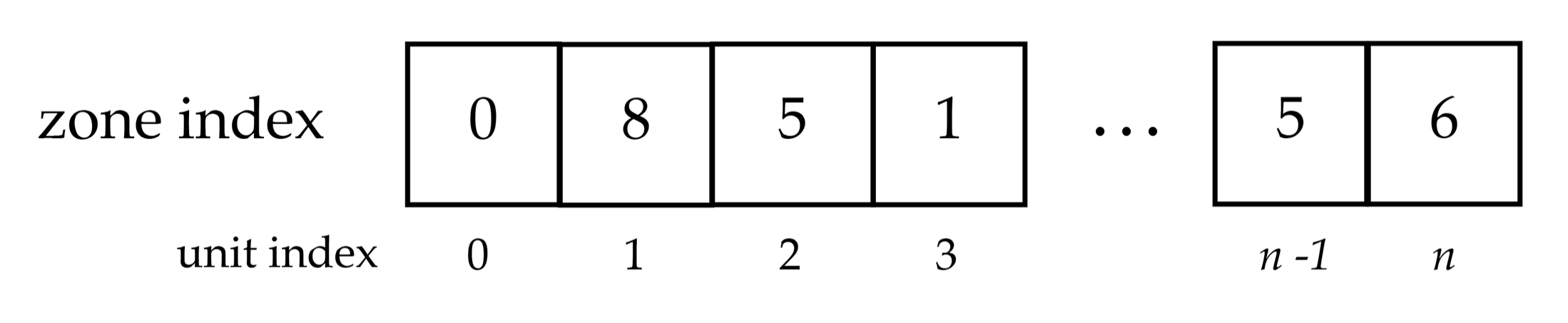}
\caption{Chromosome encoding for spatial partitioning application}
\label{fig:chromosome}
\end{figure}

To formulate an EA for this spatial partitioning problem, we may
encode the chromosome (shown in Figure~\ref{fig:chromosome}) as an
integer array with $n$ alleles, $\beta_1, \beta_2, \dots, \beta_n$.
Each allele, $\beta_i$, takes on one value from the set, $[1, k]$.
We introduce a virtual unit 0 and its associated zone 0 to indicate 
the region border outside and surrounding the units being partitioned.  
Every spatial unit is assigned to exactly one zone.

This formulation is similar to the Real-Encoded Evolutionary Monte
Carlo~\citep{LiangWong:01} except that
$\beta_i \in [0,k] \in \mathbb{Z}$ instead of
$\beta_i \in \mathbb{R}$.  While this may seem like a reduction in the
solution space, note that in our application, the state space becomes
much more complex since the spatial constraints {\em significantly}
restrict which encodings are valid or feasible solutions and render
most of the combinatorial encodings as infeasible because they violate
the spatial constraints.  In turn, the restrictions on which encodings
are valid makes the solution space traversal especially difficult
since randomly choosing alleles to alter is then very likely to result
in a solution that violates the constraints.  Movement in the solution
space thus must be carefully considered and spatially aware to avoid
wasted computational effort.

\subsubsection{Spatial Mutation Operator}

For this spatial partitioning optimization problem,~\citet{LiuCho:20}
formulated two EA spatial operators.  The first is an ejection
chain-based mutation operator, ECMUT, which selects a chromosome at
random and mutates a small number of its ``mutable'' alleles where an
allele is defined as mutable if the change in the allele does not
result in an infeasible solution.  When a small number of alleles are
mutated, at least one of the alleles is on a zone border.  This
process may be described as follows.
\begin{enumerate}
\item Randomly choose one chromosome, $\xx_i$ from the current
  population, $\mathbf{\Set}$.
\item Add a random vector subject to constraints, $\mathbf{e}$, to
  create a new chromosome $\yy_i = \xx_i + \mathbf{e}$, where
  $\mathbf{e}$ is chosen so that $\yy_i$ is a feasible solution.
\end{enumerate}

\subsubsection{Spatial Path-Relinking Crossover Operator}

The second EA operator is spatial path-relinking crossover operator
(PRCRX) that adapts the general non-spatial path relinking method to
the spatial context.  It does so by providing an ordered way to
explore and perform recombination in the neighborhood space defined by
two chromosomes~\citep{Glover:94, Gloveretal:00} that respects spatial
constraints.  Details of their operators are provided
in~\citet{LiuCho:20}.  Here, we provide just a general description.

In particular, their spatial path-relinking crossover can be described
as follows.  At step $i$,
\begin{enumerate}
\item randomly choose two chromosomes, a source solution, $\xx_s$, and
  a target solution, $\xx_t$, $s \neq t$, from the population
  $\mathbf{\Set}^{(i)}$.
\item The relinking process is comprised of a ``walk'' in the spatial
  neighborhood space between the two solutions.  Each step in the path
  converts a random allele from its value in $\xx_s$ to its value in
  $\xx_t$.  It is possible that some of set of these steps will lead
  to infeasible solutions, though spatial contiguity is always
  maintained.
\item If feasible solutions are found on the path, randomly pick
  one, $\yy$, to replace chromosome $\xx_s$ with some probability, 
  yielding a new population, $\mathbf{\Set}^{(i+1)}$.
\end{enumerate}
The ECMUT spatial mutation and PRCRX spatial crossover operators have
been shown to be effective and efficient for a large spatial
partitioning optimization application.

\begin{figure}
\centering
\includegraphics[width=5.5in]{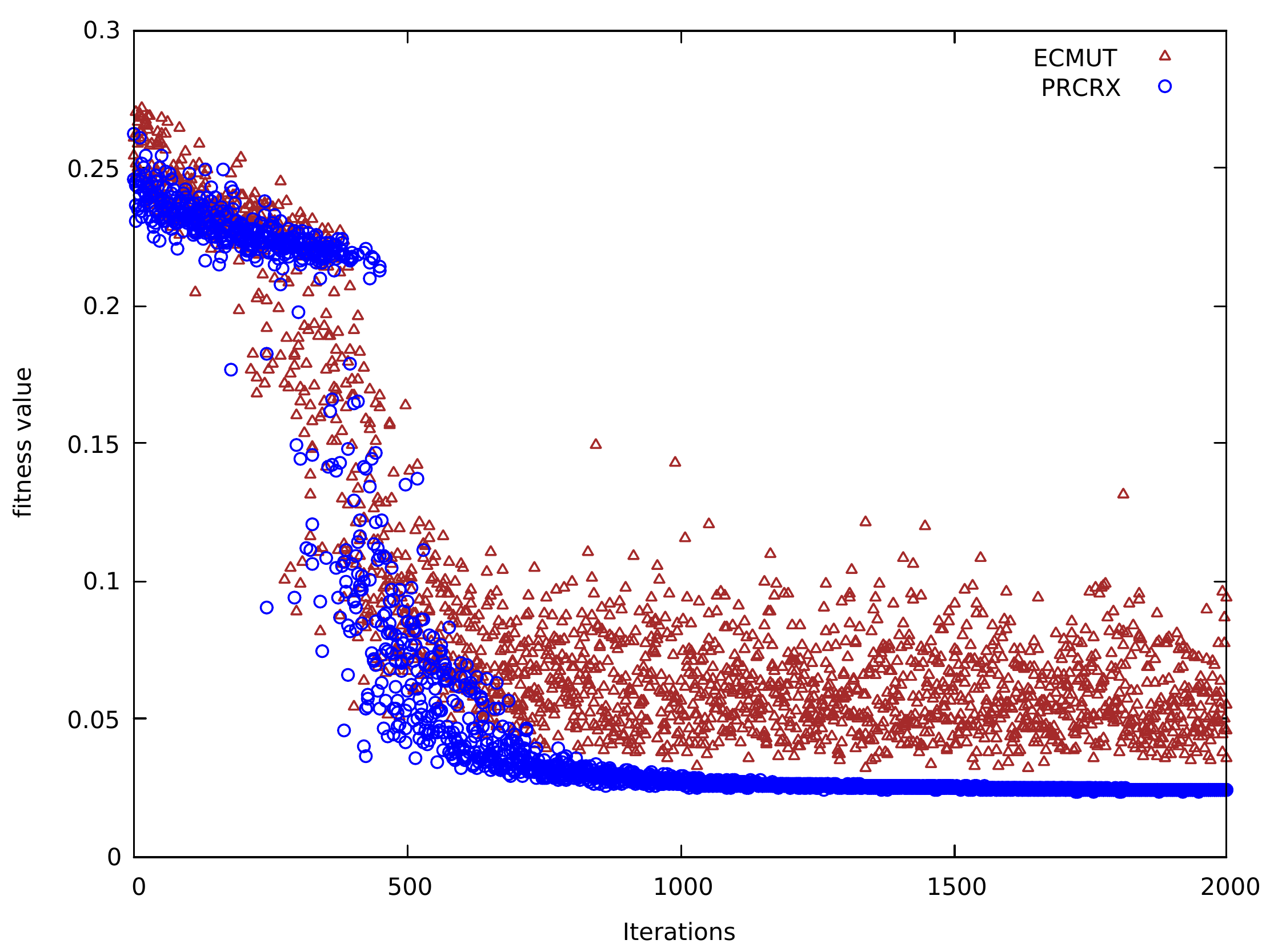}
\caption{Performance of the Spatial Path Relinking Crossover Operator (PRCRX)}
\label{fig:crossover}
\end{figure}

Figure~\ref{fig:crossover} shows the performance of these EA operators
in an optimization context.  We can see from the figure that while
ECMUT is also effective in identifying solutions with increasingly
optimal values, the spatial path relinking operator provides
significant additional improvement in the search process.  In
particular, PRCRX was able to reach lower fitness values more quickly
and more deliberately than the ECMUT operator alone, which exhibited
improved, but more variable, behavior.

In~\citet{LiuCho:20}'s empirical evaluation, they compared the
performance of their EA operators with the performance of four
previously proposed spatial optimizers and provided compelling
evidence that for the spatial partitioning optimization problem, the
ECMUT and PRCRX operators are more effective in guiding movement in
promising search directions and thus identifying better solutions.
Indeed, the idea of path relinking was originally developed to provide
guided intelligence for seeking integer solutions along the path
between two relaxed real-number solutions.

\section{Evolutionary Spatial Sampling}

Spatial optimization and spatial sampling share a common component,
the need for effective and efficient traversal of a spatial state
space.  While operators that have been designed for spatial
optimization cannot be used directly for sampling from spatial state
spaces, they can be adapted to an MCMC framework either through a
Metropolis-Hastings framework, for instance, or by providing a {\em
  proposal set} for directional sampling according to the structure of
the Multiple-Try Metropolis MCMC model.  We now discuss how this
adaptation might proceed for these particular EA operators in the
spatial partitioning problem.

\subsection{ECMUT Mutation}
\label{subsec:ecmut}

ECMUT is a fairly small and local movement, simply exchanging some
small number, $p$, of mutable units.  The adaptation of the ECMUT
mutation operator to define the transition of a Markov chain for a
standard Metropolis-Hastings MCMC is theoretically straightforward.
In particular, an ECMUT proposal is accepted with probability,
$\min(1, r)$, defined by the Metropolis-Hastings rule, where
 \begin{equation}
   r = \frac{\pi(\yy) \, T(\yy,\xx)}{\pi(\xx) \, T(\xx , \yy)} = \exp
   \{ -[H(\yy) - H(\xx)] \} \frac{T(\yy , \xx)}{T(\xx , \yy)} ,
\label{eq:rmut}
\end{equation}
and $T(\xx,\yy)$ is a proposal transition function.  When the
transition function, $T(\cdot)$, is symmetric (as it is for non-spatial
state spaces, where every allele is subject to possible mutation and
may be mutated to any other allele), only the fitness values are
needed to compute the Metropolis-Hastings ratio.

For our application, the transition function is not symmetric because
we constrain ECMUT movement to feasible alleles, at least one of which
is located on a zone border.  Hence, for the spatial partitioning
problem, we calculate the MH ratio as
\begin{equation}
r_{\tiny{\mbox{\sl MUT}}} = \frac{M_x}{M_y} \exp \{ -[H(\yy) - H(\xx)] \} ,
\label{eq:spMH}
\end{equation}
where $M_x$ is the number of mutable units that may be chosen for the
mutation step that originates from solution $\xx$ and that results in
another feasible solution.  Similarly, $M_y$ is the number of mutable
units that may be chosen for the mutation step that originates from
solution $\yy$ and that results in another feasible solution.
$H(\cdot)$ is the fitness function value for a particular solution.

To distinguish the use of the operator for differing numbers of
mutated units, we call the operator ECMUT-$p$, where $p$ is the number
of units that are mutated.  Variations of ECMUT-1 as an MCMC
transition proposal for spatial partitioning have been discussed by
others~\citep{Bangiaetal:17, MattinglyVaughn:14}.  For ECMUT-1, $M_x$
and $M_y$ are fairly straightforward to compute because they can be
enumerated with modest computational effort even for large spatial
partitioning problems~\citep{Liuetal:16}.

In particular, to compute $M_s$, where $\mathbf{s}$ is some solution,
we can assess whether each boundary unit for each zone is mutable,
which means that it can be reassigned to a neighboring zone
without disconnecting its current zone and would result in a feasible
solution after mutation. Notice that, for a planar graph, where the
edges indicate unit connectivity, the computing cost for $M_s$ depends
on the number of neighbors that must be checked for contiguity for
each unit on a zone boundary and the total number of zone boundary
units.  Because of the sparsity of a planar graph, the average cost of
checking the direct neighbors of a zone boundary unit for mutability
is $O(1)$.  In addition, following the Euler characteristic, for any
convex polyhedron surface, $V-E+F=2$, where $V$ is the number of
vertices, $E$ is the number of edges, and $F$ is the number of faces.
In our application, $V = n$, the number of units, $E$ is the number of
edges or adjacency links, and $F$ is the number of faces in our
resulting finite and connected planar graph.  We know that
$E \leq 3n - 6$ when $n \geq 3$ since any face is bounded by at least
three edges, and each edge touches two faces.  Therefore, for any
unit, the average number of direct neighbors is less than three.

The relationship between $b$, the number of boundary units for all $k$
zones, and $n$ can be modeled as the total length of the zone
perimeters and the sum of the zone areas.  We can make an analogy to
cutting a pizza into $k$ pieces where the shape of each piece is
inconsequential as long as the $k$ pieces collectively encompass the
whole pizza.  Here, each pizza piece is akin to a zone.
The upper bound of $b$ is $n$, regardless of the value of $k$.  For
example, consider an $n = q \times k$ grid, when each row of $q$ cells
is a zone in a $k$-partition solution.  The lower bound of $b$ is $k$.
This would occur in the scenario where one zone has $(n-k+1)$
contiguous units while the $(k-1)$ remaining zones contain one unit
from the rest of the $(k-1)$ units.  Hence, we know that
$\frac{k}{n} \leq \frac{b}{n} \leq 1$.  An important observation is
that, when $k$ is fixed, the rate of decrease for the ratio,
$\frac{b}{n}$, is faster than the rate of increase for $n$.  That is,
while the area increases linearly with $n$, the perimeter increases in
proportion to $\sqrt{n}$.  The manifestation in our problem is that,
as the problem size increases, the cost of computing $M_s$ rises much
slower than the linear increase of $n$.
 
While ECMUT-1 provides a proper Markov transition with a computationally
feasible MH ratio, we have already discussed the issues associated 
with the ECMUT-1 transition.  First, while this type of transition
in a Metropolis-Hastings MCMC is simple conceptually and
operationally, and results in a fluid Markov chain, the performance of
such an operationalization is inefficient and possibly insufficient if
the state space is particularly large or complex.  Unless the
performance can be greatly improved, perhaps via massive computational
resources, this type of transition proposal will be inadequate for
large scale applications.  Second, even for a relatively small spatial
partitioning application, because the spatial constraints render large
numbers of the combinatorial encodings as infeasible, this Markov chain
is likely to become trapped in localized regions that are disconnected
from other feasible solutions, resulting in a reducible chain that is
incapable of sampling from the global state space.

An ECMUT-$p$ operator, where $p > 1$, produces larger transitions,
which contributes to searching efficiency and reduces (though do not
solve) the issues associated with disconnected state space regions
since invalid encodings can be traversed between the first and the
$p$th single unit step.  If we are able to enumerate the set of
possible transitions of size $p$ from a solution $\xx$, we can compute
the MH ratio (\ref{eq:spMH}) directly.  The tradeoff is, as $p$
increases, this combinatorial calculation becomes increasingly
computationally expensive, requiring innovations in algorithmic
sophistication.\footnote{\citet{Fifieldetal:19} bypass the need for
  computational advancements by proposing an approximation for the
  transition probability, $p! \left( \frac{1}{u_x}\right)^p$.
  However, their approximation is based on intuition and was not
  theoretically derived.  They acknowledge that ``it is difficult to
  develop a rigorous theoretical justification for the proposed
  approximation.''  In addition, while they provide an empirical
  example, the closeness of this approximation to the true value is
  unknown in general and has not been rigorously tested or evaluated.}

\subsection{Crossover and Recombination}

EA crossover operators have been successfully integrated into an MCMC
framework.  For example,~\citet{LiangWong:00} adapt a binary-encoded
EA, where the alleles, $\beta_i \in \{ 0, 1\}$, produce a chromosome
that is a binary vector.  Their crossover operators include a 1-point,
2-point, and uniform crossover.  They choose the first parent
chromosome, $x_i$ from the population, $\xx$, using a roulette wheel.
The second parent, $x_j$, is chosen randomly from the remaining
chromosomes.  From the two parent chromosomes, two offspring
chromosomes, $y_i$ and $y_j$, are generated.  Without loss of
generality, assume $H(x_i) \ge H(x_j)$ and $H(y_i) \ge H(y_j)$.  A new
population,
$\mathbf{y} = \{ x_1, \dots, y_i, \dots, y_j, \dots, x_n \}$, is
proposed and is accepted with probability, $\min(1, r_c)$, where
\begin{equation}
  r_c = \frac{\pi(\mathbf{y}) \, T(\yy , \xx)}{\pi(\mathbf{x}) \, T(\xx
    , \yy)} = \exp \{ -[H(y_i) - H(x_i)] / t_i - [H(y_j) - H(x_j)] /
  t_j\} \frac{T(\yy , \xx)}{T(\xx , \yy)} 
\label{eq:bcrossover}
\end{equation}
is the Metropolis-Hastings ratio, and
T$(\xx , \yy) = P((x_i, x_j) \, | \, \mathbf{x}) \, P((y_i, y_j) \, |
\, (x_i, x_j))$.  Within the transition probability,
$P((x_i, x_j) \, | \, \mathbf{x})$, the probability of selecting
$(x_i, x_j)$ from the population $\mathbf{x}$, is
\begin{equation}
P((x_i, x_j) \, | \, \mathbf{x}) = \frac{1}{(n-1)Z(\mathbf{x})}
[\exp [ -H(x_i) / t] + \exp \{ -H(x_j) / t\}], 
\end{equation}
where $Z(\mathbf{x}) = \sum_{i=1}^n \exp \{ -H(x_i) / t\}$.
Since
this crossover operator is symmetric, the ratio of transition
probabilities reduces to 1, and so only the selection probabilities,
which are either chosen randomly or based on fitness values, are
needed to determine the acceptance probability.

Real-encoded EAs can be more intimidating than binary-encoded EAs in
the sense that the non-spatial real-encoded model,
$x = \{ \beta_1, \beta_2, \dots, \beta_k\}$, with
$\beta_i \in \mathbb{R}$, encompasses a much larger solution space to
traverse.  However, when moving to the non-spatial real-encoded values
no additional complications are introduced in the adaptation for
symmetric transition probabilities.  This occurs for simple crossover
operators like the $k$-point or uniform crossovers.

Unfortunately, as we have discussed, these simple crossover operators
are ineffective for spatial applications because they are spatially
unaware.  Indeed, this was the impetus for deriving new EA operators
for spatial optimization where we designed the spatially cognizant
PRCRX operator to produce both larger movements and enable a more
diversified search.  Our empirical evaluation showed that PRCRX
offered significant improvement in both the effectiveness and the
efficiency of the optimization search process~\citep{LiuCho:20},
making it an auspicious prospect for an MCMC transition.  Adaptation
of PRCRX into an MCMC context, however, is not as simply realized as
it is for the common and basic binary EA crossover operators.

\subsection{The Spatial Path Relinking Crossover}
\label{subsec:prcrx}

We adapt PRCRX, the spatial path relinking crossover operator that we
developed as part of an optimization heuristic, to an MCMC algorithm
through the Multiple-Try Model framework.  Here, the purpose of the
optimization crossover operator is to create a Markov transition {\em
  proposal set}.  One proposal state from this set is chosen
probabilistically and is accepted or rejected according to a
generalized Metropolis-Hastings ratio.

We begin with a set of $q$ parallel chains,
$\XX_1,\XX_2, \dots, \XX_q$, that are randomly generated.  Their state
at time $t$ is $\xx_1^{(t)}, \xx_2^{(t)}, \dots, \xx_q^{(t)}$,
respectively.  For the spatial path relinking crossover, we need a
current state and a target state.  If, at time $t$, for chain $\XX_i$,
we want to invoke a spatial path relinking crossover transition, then
state $\xx_i^{(t)}$ becomes the current state.  We randomly select
another chain, $\XX_j, j \neq i$ from the set or population of all
other Markov chains.  The current state of that randomly chosen chain,
$\xx_j^{(t)}$, becomes the target state.\footnote{The MTM framework
  requires $T(\xx,\yy) > 0$ if and only if $T(\yy,\xx) > 0$.  That is,
  the transition to $\yy$ from $\xx$ is possible if and only if this
  transition is reversible with non-zero probability.  To ensure that
  $\xx$ could be chosen as the anchor solution, whenever a path
  relinking transition is accepted, we enlarge the target pool by one
  by inserting the current solution, $\xx_i^{(t)}$, into the target
  pool.  Since we just traversed a path that connects $\xx$ to $\yy$,
  we know such a path exists.  Hence, if $\yy$ becomes the current
  state, to satisfy the reversibility requirement, it only has to be
  possible that $\xx_i^{(t)}$ could be chosen as the target solution.}
~\citet{RobertsGilks:94} established that essentially any way of
choosing the target state, $\xx_j^{(t)}$, is appropriate provided that
$\xx_i^{(t)}$ and $\xx_j^{(t)}$ are independent.

\begin{figure}
\centering
\includegraphics[width=6.5in]{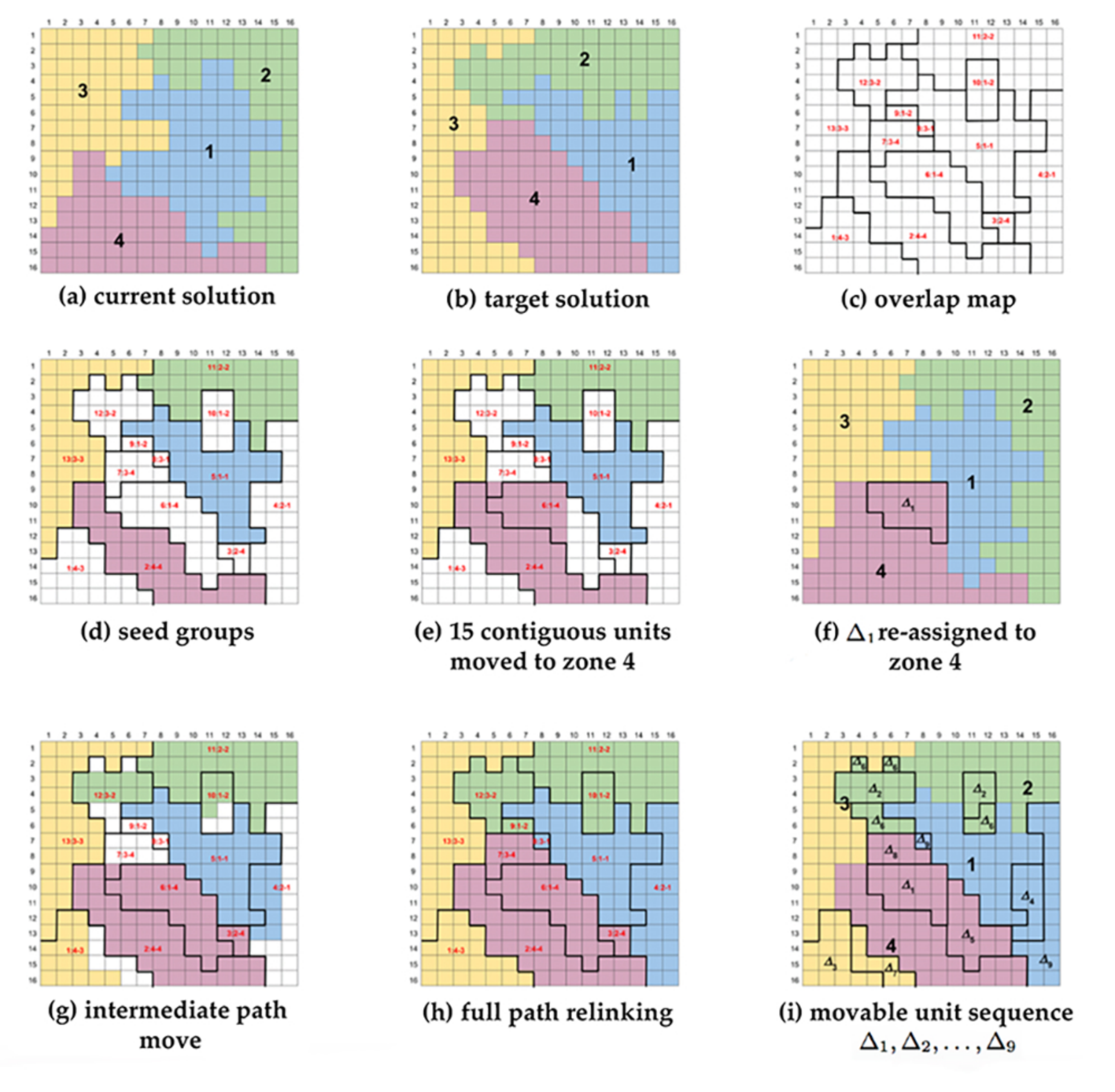}
\caption{Illustration of the Crossover Operator (PRCRX)}
\label{fig:mcmccrossover}
\end{figure}

The spatial path relinking crossover, PRCRX, begins at state
$\xx_i^{(t)}$ and moves in the direction of state $\xx_j^{(t)}$.  Each
step in the crossover heuristic involves the swap of some set of
contiguous units from their zone in the current state to their zone in
the target state.  Figure~\ref{fig:mcmccrossover} illustrates this
process.  The current state is shown in subfigure (a).  The target
state is shown in subfigure (b).  Each of the outlined regions in
subfigure (c) has a unique source/target zone label, indicating its
zone assignment in the source state and its zone assignment in the
target state.  Since there are 4 zones, there are $4*4=16$ possible
unique current/target zone labels, though all of these zone labels
will not necessarily be in an overlap.  In the illustration, only 13
of the 16 possible groupings appear.

The path relinking begins with a set of $k$ seed groups, where $k$ is
the number of zones in the partitioning problem.  These seed groups
can be chosen in any way as long as the resulting groups each have
unique target zones.  Subfigure (d) shows one way in which to choose
these seed groups, i.e., by picking those groups where the current and
target zone labels are the same.

Figure~\ref{fig:mcmccrossover} shows one possible path that traverses
the entire distance from the source state to the target state.  Note
that full traversal of the path from the current state to the target
state is not necessary since the purpose is simply to explore the
feasible intermediate states in the neighborhood space, not
necessarily to complete the path.  Subfigure (f) shows one
intermediate state along one path that results from the re-assignment
of the region labeled $\Delta_1$.  Many intermediate/proposal states
exist along the various possible paths in the spatial neighborhood
space.

Let $C$ be the total number of sets of contiguous units that need to
be moved or re-assigned to complete the path.  A proposal state is the
result of the swap of some number, $c$, of these components,
$1 < c < C$.  The distance between states $\xx_i^{(t)}$ and
$\xx_j^{(t)}$ is defined as $d = \sum^C_{z=1}{|G_z|}$, where $z$ is a
zone identifier, $G_z$ is a set of contiguous units, $C$ is the number
of sets, $G_z$, in the source zone that can potentially be moved to
the target zone, and $|G_z|$ is the cardinality of the set of
contiguous units in the group $G_z$.  At most, $d$ units, contained
within $C$ connected components need to be moved to complete the
entire path from the current state to the target state.

The direction of the movement from the current state is determined by
the target state.  The magnitude of the movement in this direction is
chosen randomly by sampling
$d_m \sim \mbox{Unif}[ \{ |G_z|_{\forall z} \}]$.  To identify a set
of $m$ proposal states, $\yy_1, \yy_2, \dots, \yy_m$, at various
distances, $d_m$, along the path in the spatial neighborhood space of
state $\xx_i^{(t)}$ and state $\xx_j^{(t)}$, we draw $m$ samples.

Once we have our proposal set, $\YY = \{ \yy_1, \yy_2, \dots, \yy_m\}$,
we choose a state, $\yy$, from this set with probability proportional
to $\xi (\xx, \yy) = \pi(\xx) \, T(\xx, \yy) \, \lambda(\xx, \yy)$, where
$\lambda(\xx, \yy)$ is a non-negative symmetric function and
$T(\xx,\yy)$ is the proposal transition function defined by the path
relinking operator.  Note that $T(\xx,\yy)$ does not need to be
symmetric.  In addition, the only requirement for $\lambda(\xx, \yy)$
is that $\lambda(\xx, \yy) > 0$ whenever $T(\xx, \yy) > 0$.  How to
identify an optimal $\lambda(\xx, \yy)$ for a particular application
is an open question.  However, \citet{Liuetal:00}, in their set of
numerical experiments, found that the performance of an MTM sampler
was insensitive to the choice of $\lambda(\xx,\yy)$ and that the
simplest choice is $\lambda(\xx,\yy) \equiv 1$.  This is the choice we
utilize.

If $c$ connected components is the number of connected components that
have been moved from the current state $\xx$ to the target state,
$\yy$, then we can see that $T(\xx, \yy) = C (C-1) \dots (C-(c-1))$
and then $\xi (\xx, \yy) = T(\xx, \yy) \exp \{H(\yy) \}$.

Once the state $\yy$ is chosen, we generate a second set,
$\{ \xx^* \} = \{ \xx_1^*, \xx_2^*, \dots, \xx_{m-1}^* \}$, from
$T(\yy, \cdot)$, which is a transition that is also defined by the
spatial path relinking model, but $\yy$ is now the current state,
$\xx_i^{(t)}$ is the target state, and $\{ \xx^* \}$ is a set of
intermediate states that lie in the neighborhood space between $\yy$
and $\xx_i^{(t)}$.  We then accept $\yy$ as the next Markov chain
state with probability defined by the generalized Metropolis-Hastings
ratio,
\begin{equation}
r_{\tiny{\mbox{{\sl PRCRX}}}} = \min \left\{ 1, \frac{\xi (\yy,\xx) + \sum_{j=1}^{m-1} \xi (\yy_j,
      \xx)}{\xi (\xx,\yy) + \sum_{j=1}^{m-1} \xi (\xx_j^*, \yy)} \right\} ,
\end{equation}
and let $\xx_i^{(t+1)} = \xx_i^{(t)}$ with the remaining probability.

This process allows us to update one Markov chain, $\XX_i$, with
information from another Markov chain, $\XX_j$, that is randomly
chosen to be the target chain.  The next step, $\xx_i^{(t+1)}$, for
the current Markov chain, $\XX_i$, is updated with an MTM transition
along the direction defined by the path between $\xx_i^{(t)}$ and
$\xx_j^{(t)}$.

\subsection{Evolutionary Markov Chain Monte Carlo Algorithm (EMCMC)}

Our Evolutionary Markov Chain Monte Carlo (EMCMC) algorithm combines
the advantages of evolutionary algorithms as optimization heuristics
for state space traversal and the theoretical convergence properties
of Markov Chain Monte Carlo algorithms for sampling from unknown
distributions.  We encompass these two algorithms within the framework
of a Multiple-Try Metropolis Markov Chain model that incorporates a
generalized Metropolis-Hasting ratio.  The general framework is as
follows.

\begin{enumerate}
\item Initialize $q$ parallel chains, $\XX_1, \XX_2, \dots, \XX_q$ at
  separate randomly generated feasible solutions,
  $\xx_1^{(0)}, \xx_2^{(0)}, \dots, \xx_q^{(0)}$.
\item For each iteration of the algorithm, $t = 0, 1, \dots, T$, the
  current state of each chain, $\xx_i^{(t)}$, is updated to
  $\xx_i^{(t+1)}$.  The state at iteration $t+1$ is determined with
  either a mutation transition, with probability $p_m$, or a crossover
  transition, with probability $p_c = 1-p_m$.  The values of $p_m$ and
  $p_c$ may be the same or different across the $n$ chains.
\item If the next step is determined with a mutation transition,
  generate a proposal transition via the ECMUT operator, and accept or
  reject that proposal with MH ratio,
\begin{equation*}
  r_{\tiny{\mbox{\sl MUT}}} = \frac{M_x}{M_y} \exp \{ -[H(\yy) - H(\xx)] \} .
\end{equation*}
\item If the next step is a proposal from the PRCRX spatial path
  relinking crossover operator, accept the proposal with the
  generalized Metropolis ratio,
\begin{equation*}
r_{\tiny{\mbox{{\sl PRCRX}}}} = \min \left\{ 1, \frac{\xi (\yy,\xx) + \sum_{j=1}^{m-1} \xi (\yy_j,
      \xx)}{\xi (\xx,\yy) + \sum_{j=1}^{m-1} \xi (\xx_j^*, \yy)}
  \right\}  .
\end{equation*}
\end{enumerate}

By adapting the ECMUT mutation operator and the spatial PRCRX
crossover operators into the MCMC framework as described in
Sections~\ref{subsec:ecmut}--\ref{subsec:prcrx}, we arrive at a new
MCMC algorithm, which we term an Evolutionary Markov Chain Monte Carlo
algorithm (EMCMC), for sampling constrained spatial partitions.  EMCMC
utilizes optimization heuristics to guide the movement of Markov
chains.  This enables large movements in the state space while
supplying a way for these large movements to identify states that are
not unlikely to be associated with a reasonably large
Metropolis-Hastings ratio.  The Multiple-Try Method provides a
generalized Metropolis-Hastings ratio that enables the integration of
optimization methods with MCMC algorithms that is able to sampling
spatial partition with constraints from spatial state spaces.

\section{Empirical Example}

\begin{figure}[H]
\centering
\includegraphics[width=3.5in]{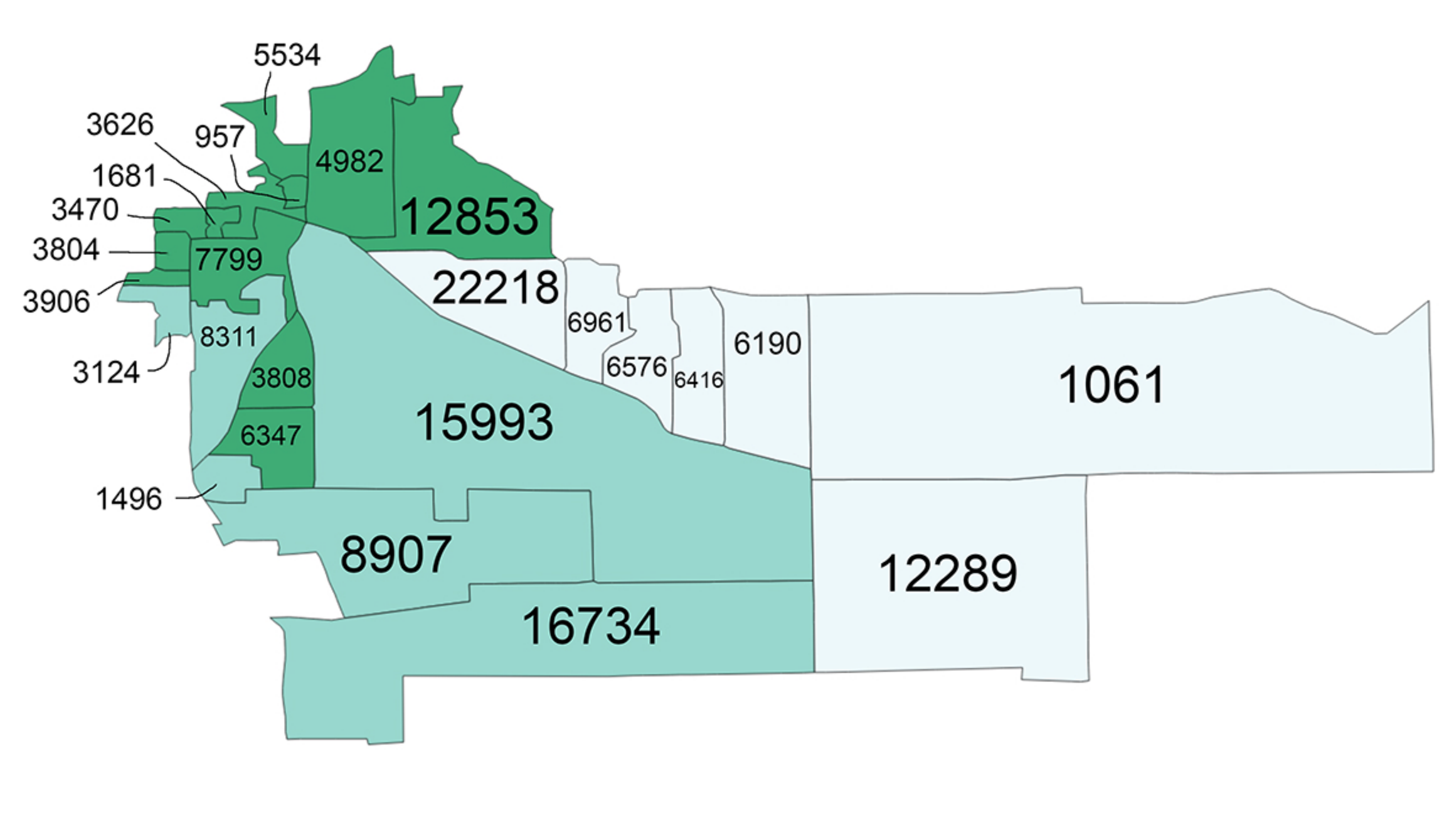}
\hspace{8mm}
\includegraphics[width=2in]{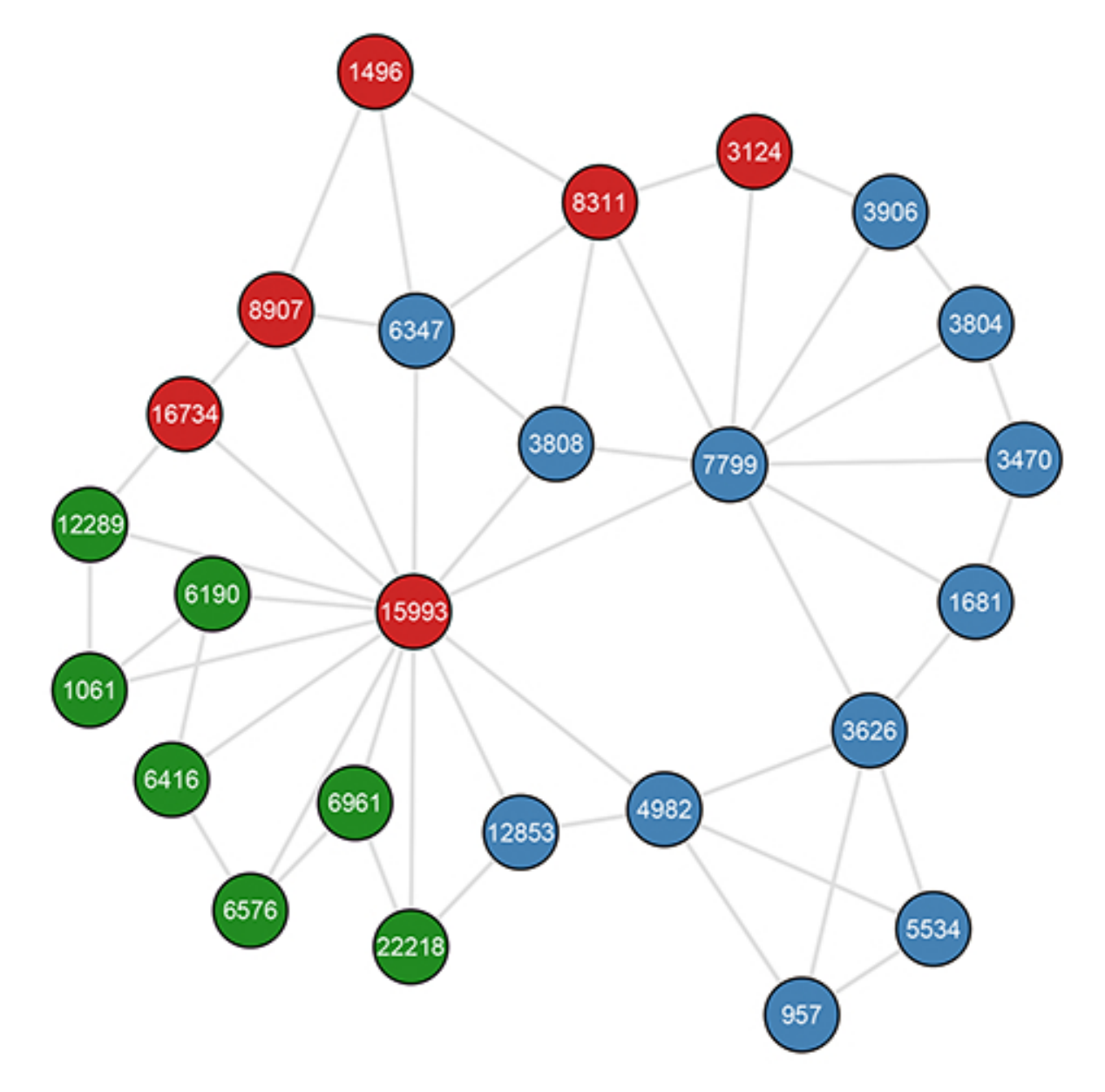}
\caption{Spatial Partitioning Example.  A representation with spatial
  geographic units is shown on the left.  A graph representation is
  shown on the right.}
\label{fig:map1}
\end{figure}

 We demonstrate the properties and effectiveness of EMCMC with an
empirical application.  In this application, there are 25 spatial
units that we seek to partition into 3 spatially contiguous and
separate zones.  The plot on the left in Figure~\ref{fig:map1} shows
these units and their spatial configuration as a collection of spatial
geographic units.  This representation is not particularly conducive
to computational structures, and so we transform these units into a
graph theory format.  The corresponding graph theory representation is
shown on the right in Figure~\ref{fig:map1}.  The edges in the graph
indicate spatial adjacency.  Each node is able to preserve the
attributes of the geographic unit via node weights/attributes.  The
unit weights are shown by the number in each of the nodes.

In this example, a feasible solution is defined as an exhaustive and
mutually exclusive partition of the 25 nodes into 3 disjoint
contiguous subgraphs such that the total weight in each subgraph is
within 10\% of the total weight in each of the other subgraphs.  The
coloring in the map and the graph displays one feasible partition of
the units into 3 zones/subgraphs that satisfy the contiguity and
weight balancing requirement.  The total weight of the red subgraph in
the graph is 54,565.  The total weight of the green subgraph is
61,711.  The total weight of the blue subgraph is 58,767.  If we
measure the weight balance as $\max_i | w_i - \mu | / \mu$, where
$\mu = \sum_i w_i / k$, and $k$ is the number of zones, then this
partition has a weight balance score of 0.058, which is below our
threshold value of 0.10.  We seek to produce a uniform sample of the
set of all such feasible partitions.

With no constraints (i.e., no contiguity requirement and no weight
balancing requirement), there are $S(25,3) = 141,197,991,025$ ways to
partition these spatial units into 3 zones, where $S(n,k)$ is a
Sterling number of the second kind.  When a contiguity requirement is
imposed, the number of feasible partitions reduces to 117,688 possible
solutions.  When the weight balancing requirement is added, only 927
of the possible partitions remain feasible partitions.  Evidently, the
constraints significantly reduce the feasible portion of the overall
state space.

\begin{figure}
\centering
\includegraphics[width=4.8in]{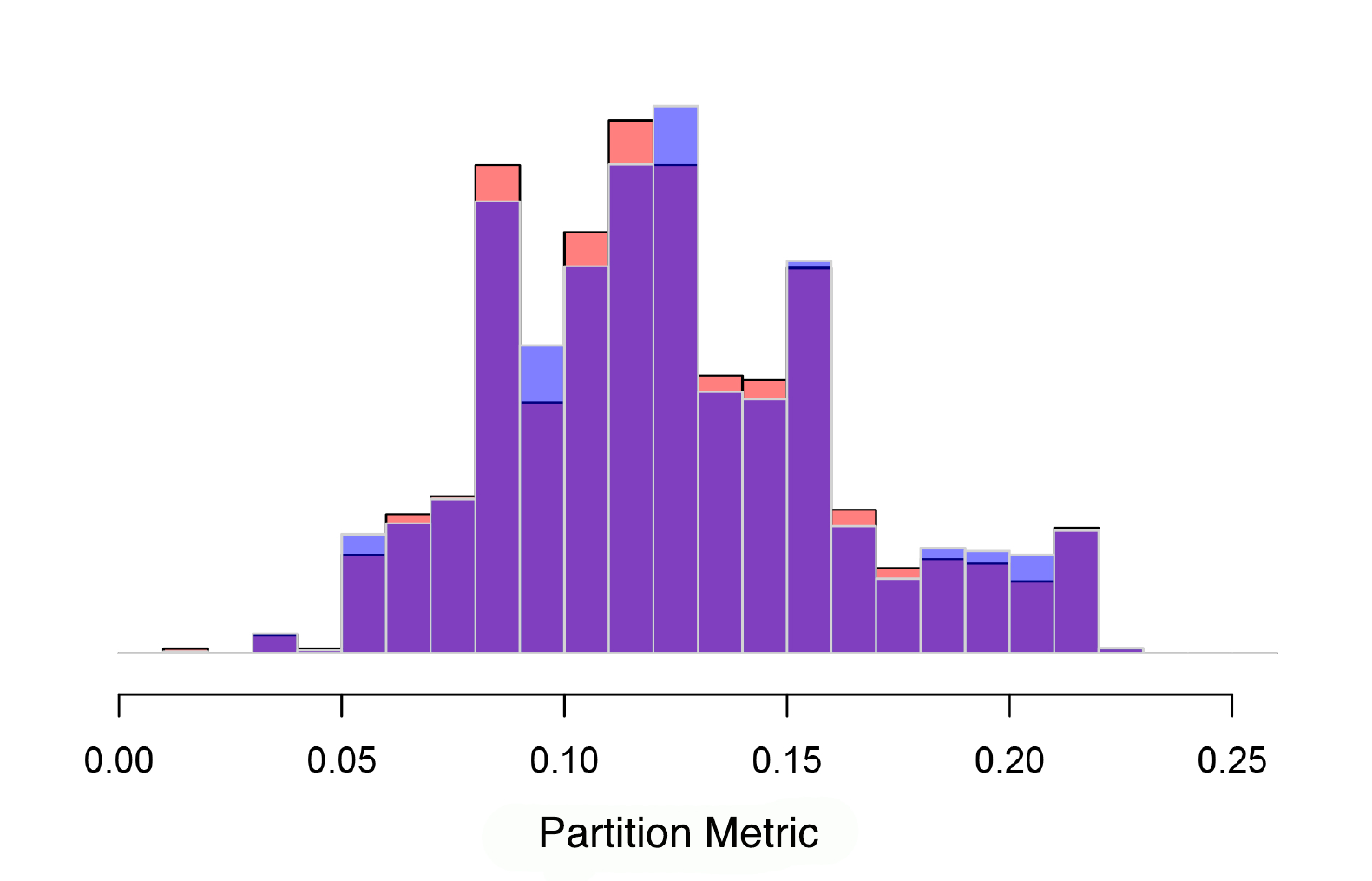}
\caption{Spatial Partitioning Application Example.  The true
  underlying distribution of the partition metric, $f(X)$, is shown
  with a red histogram.  Overlaid on that red histogram is a blue
  histogram that shows the metric for the states visited by our Markov
  chains.  In purple is the overlap of these two distributions.}
\label{fig:5sec}
\end{figure}

We employed our EMCMC algorithm with 16 parallel chains.  In 5 seconds
of time, each chain visited approximately 7,000 feasible states or
partitions where feasible is defined as satisfying the contiguity and
the weight balancing requirements.  Let $X$ represent a feasible
state.  Since these states are spatial partitions, the state space
does not have a convenient structure that is simple either to order or
to represent in ${\mathbb{R}}$.  To alleviate this problem, let
$f: X \rightarrow {\mathbb{R}}$ be a function that takes a partition,
$X$, and maps it to a metric in ${\mathbb{R}}$.  This function may be
an arbitrary function.  The partition metric we use in our empirical
example is,
\begin{equation}
  f(X) = \frac{1}{2} \sum_{i = 1}^k \frac{w_i}{\sum_{i=1}^k w_i} \frac{
    \mid r_i - R \mid}{R(1-R)} ,
  \label{eq:dissimilarity}
\end{equation}
where $i = 1, \dots\ k$ is a zone index, $w_i$ is the aggregated
weight in zone $i$, $r_i$ is the proportion of weight in zone $i$ with
a particular characteristic, $r$, and $R$ is the proportion of that
characteristic, $r$, across all zones.  The resulting distribution of
this partition metric, $f(X)$, among the visited states is shown in
Figure~\ref{fig:5sec}.  The true underlying distribution of the
partition metric is shown with a red histogram.  Overlaid on that red
histogram is a blue histogram that shows the metric for the states
visited by our Markov chains.  In purple is the overlap of these two
distributions.  As we can see, the overlap of the states visited by
the Markov chains and the true underlying distribution, while not
exact, is fairly close.  The histograms and their overlap provide
evidence of the ability of our EMCMC algorithm to uniformly sample the
underlying spatial state space.

\section{The Challenges with Scaling to Larger Applications}
\vspace{5mm}

While EMCMC was able to successfully sample the feasible spatial
partitions in this spatial state space, this particular spatial
partitioning application, while not trivial, is still small in size
and does not reach the level of complexity that accompanies both
larger problem sizes and/or additional constraints.  Hence, while this
spatial state space provides some proof of concept, additional hurdles
must be overcome in order to ensure successful navigation for more
complex applications.  While our small example does not, itself,
produce a sufficiently challenging application, it has two important
roles.  First, it is small enough that we can enumerate all of the
spatial partitions and thus allows us to test the ability of our
algorithm to produce a uniform sample from the true underlying
distribution.  Second, it is large enough to allow us to explore the
structure and root of the obstacles that manifest in larger and more
complex applications.

\subsection{Solution Sparsity}

\begin{figure}
\centering
\includegraphics[width=2.5in]{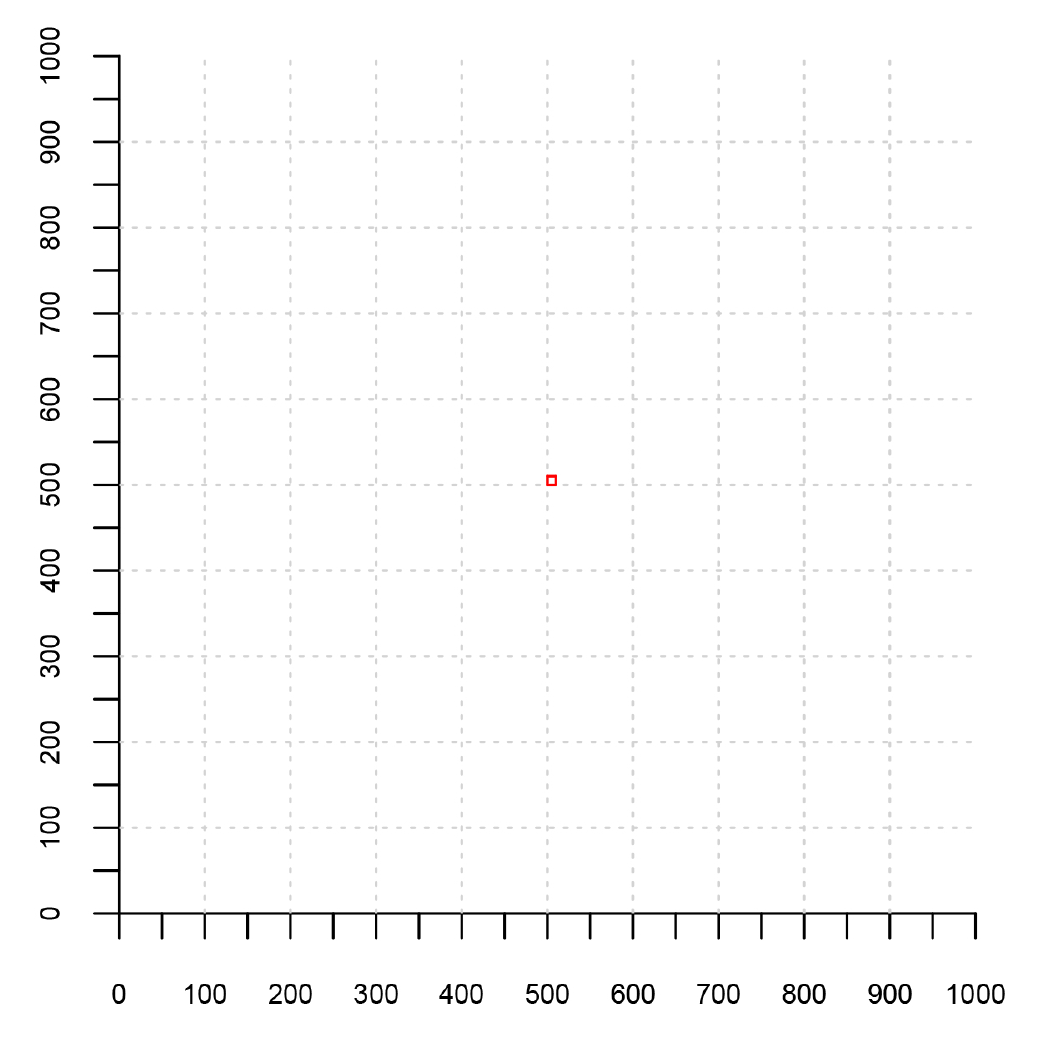}
\includegraphics[width=2.5in]{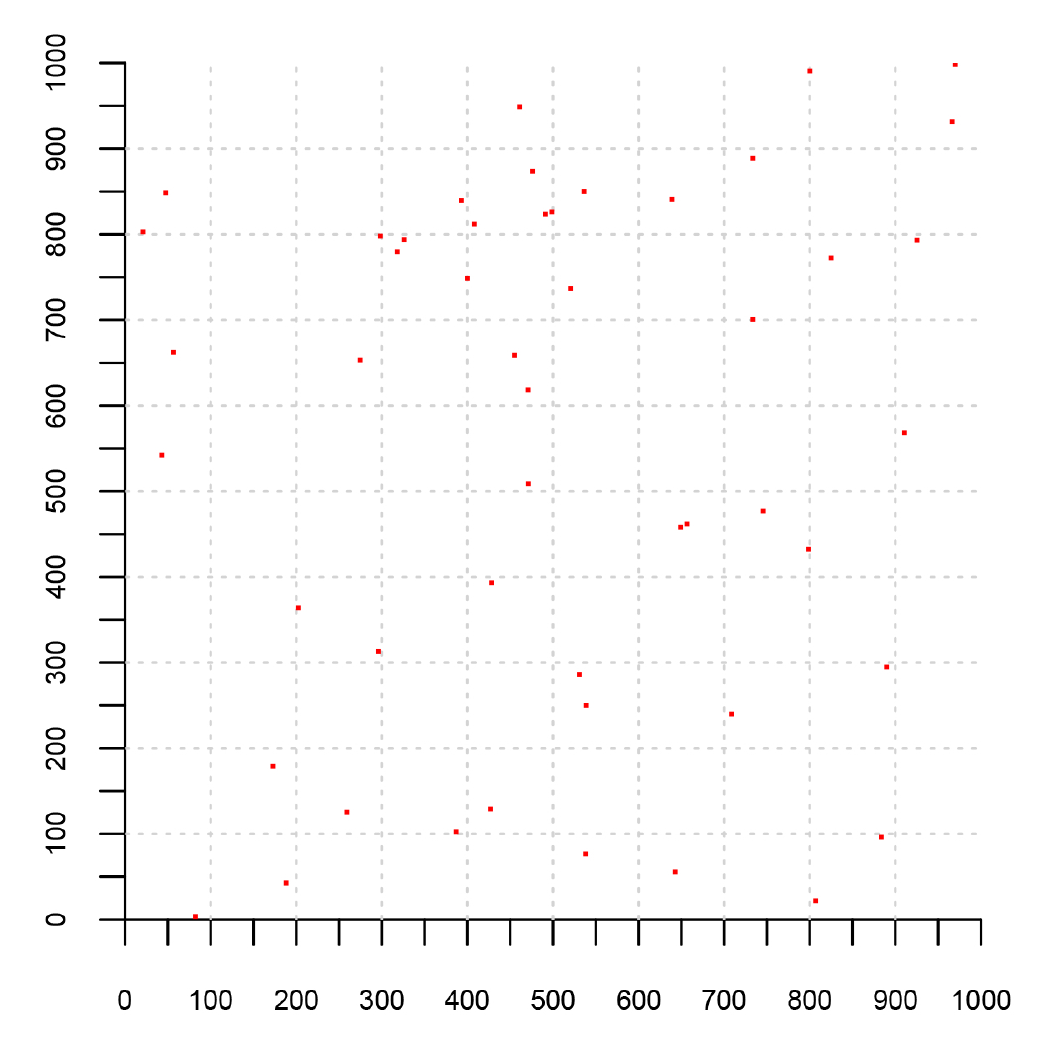}
\caption{Visualizing the state space and the portion of the state
  space that contains feasible solutions.  The set of feasible
  solutions is greatly reduced from the full set of combinatorial set
  partitions.  The individual feasible solutions are scattered about
  the solution space in an unstructured and unknown manner.}
\label{fig:size}
\end{figure}

One issue is solution sparsity.  In our small problem, the simple
unconstrained combinatorial construction of sets has more than
$10^{11}$ possible solutions.  When the spatial contiguity constraint
is imposed, the number of feasible partitions is reduced many orders
of magnitude to $10^5$.  That is, fewer than 0.0001\% of the set
partitions are spatially contiguous.  Though it is obvious that
considerations of spatially defined locality and adjacency have a
direct and explicit effect on solution feasibility, this example
highlights that even in a small sized application, the effect is
dramatic.

Figure~\ref{fig:size} helps us visualize the precipitous decline in
the number of feasible solutions from the number of unconstrained
combinatorial set partitions.  In the figure on the left, if the plot
region represents the full set of combinatorial set partitions, then
the red square in the middle is {\em 100 times larger} than the set of
contiguous partitions.  Pointedly, the feasible state space is {\em
  much} smaller than the entire decision space.

\subsection{Disconnected Solution Space}

While the red square helps us visualize the size of the feasible
space, note that the feasible solutions are not concentrated in the
central part or in {\em any} part of the entire solution space.
Instead, the feasible solutions are scattered throughout the decision
space in an unstructured and unknown manner, as illustrated in the
plot on the right in Figure~\ref{fig:size}.  While the size of the
dots is now too large (collectively, they should be 100 times smaller
than the red square in the plot on the left), they are correctly shown
as scattered throughout the space.  For any reasonably sized spatial
partitioning problem, then, an intelligent spatially guided search is
essential since simply random movement in the decision space almost
surely identifies an infeasible solution, which results in enormous
wasted computational effort~\citep{LiuCho:20}.

Importantly, note the critical implications that arise from the choice
of the graph theoretic representation.  Computationally, the graph
theoretic framework is advantageous because it permits a simple way to
incorporate spatial adjacency with convenient discrete structures and
provides an organizational framework for state space traversal.  It
also facilitates a way to define a fully connected state space such
that beginning at any contiguous state, one can reach any other
contiguous solution via a sequence of deletions/creations of single
edges between nodes.  Unsurprisingly, accounting for contiguity in
this graph traversal is relatively simple since the basis of the graph
structure is spatial contiguity.  However, perhaps also
unsurprisingly, then, graph traversal accounting for other constraints
is not so easily facilitated and requires dedicated effort and
ingenuity.

\begin{figure}
\centering
\includegraphics[width=2in]{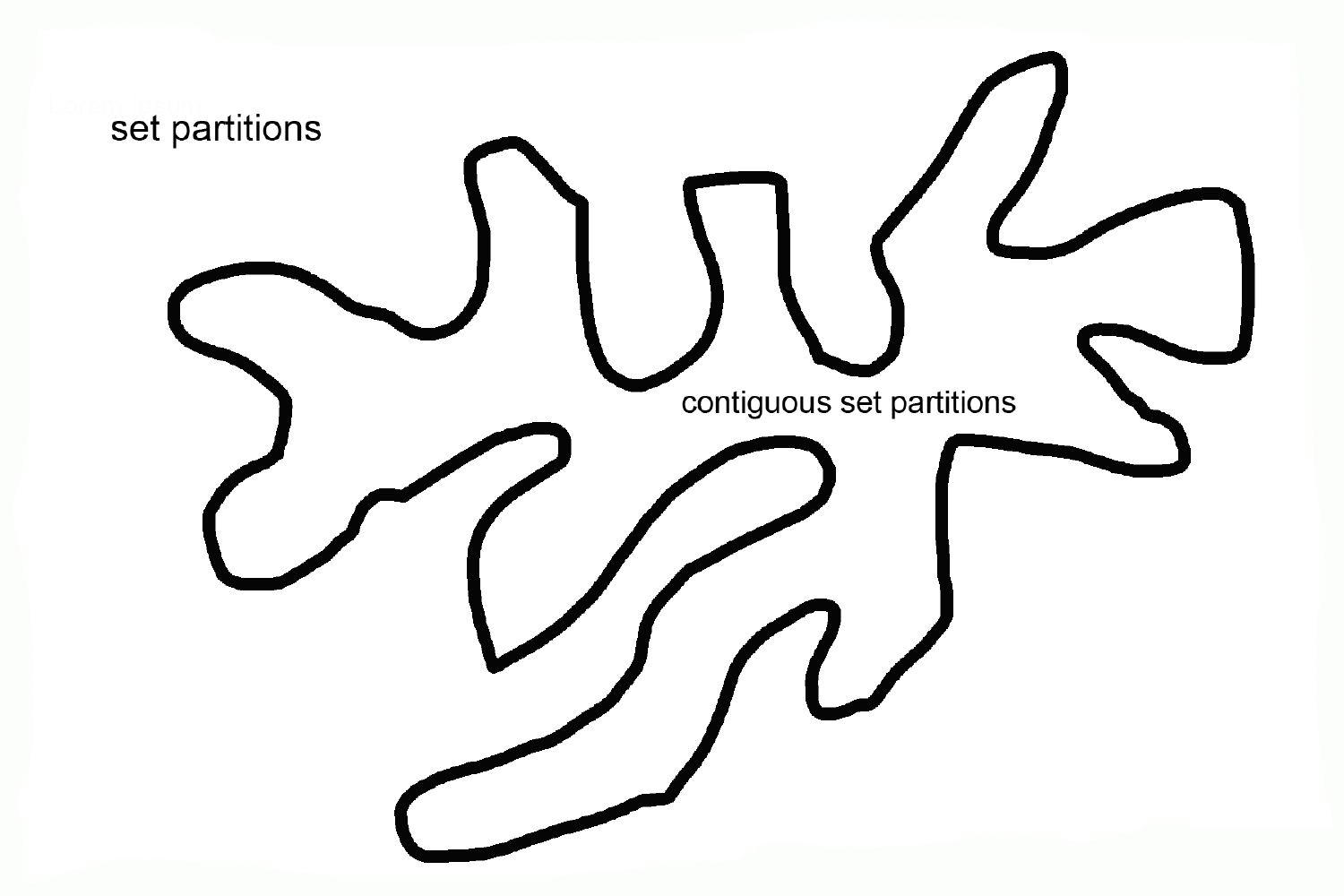}
\hspace{2cm}
\includegraphics[width=2in]{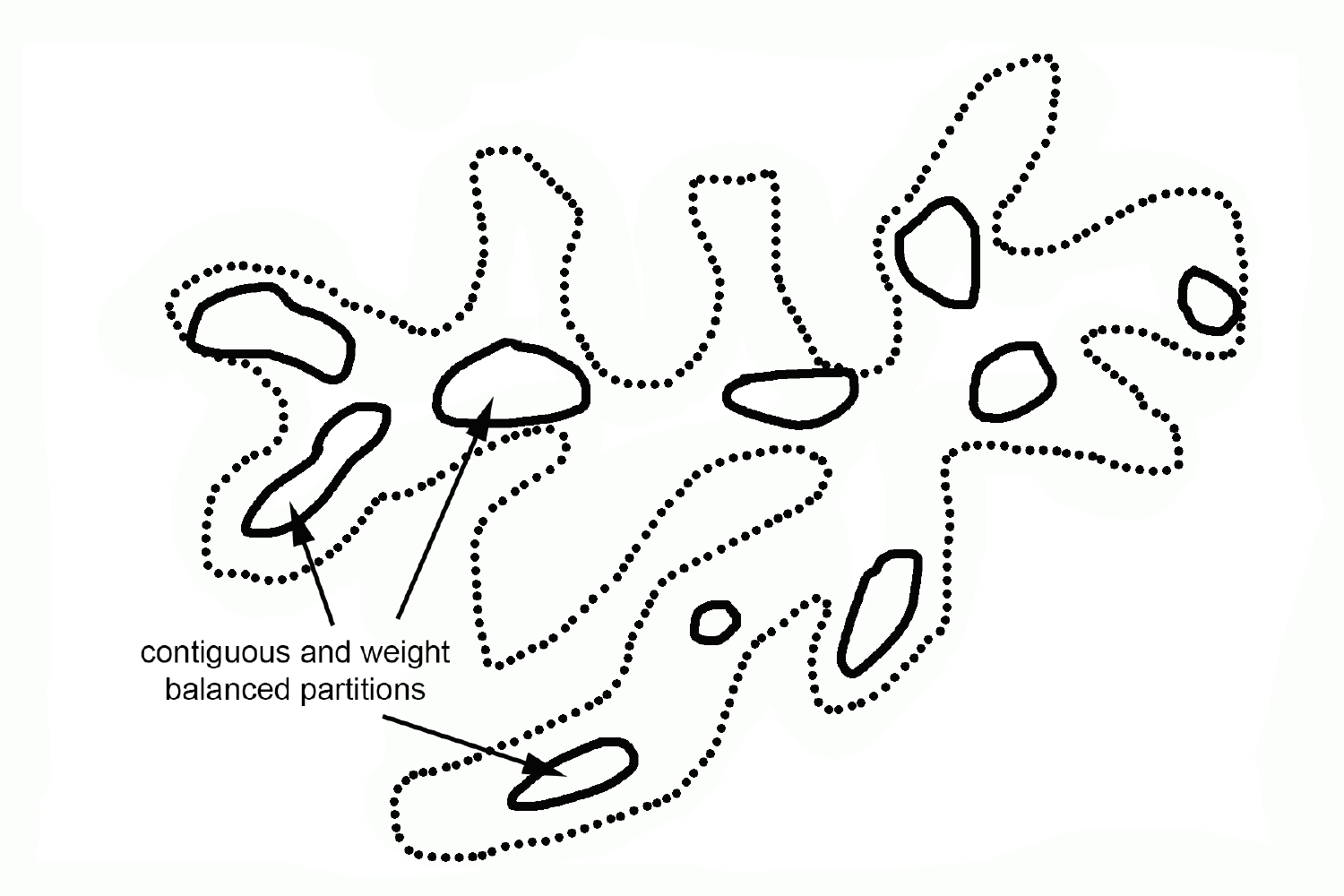}
\caption{The figure on the left illustrates a fully connected subspace
that is comprised of all solutions that satisfy the contiguity
constraint.  The figure on the right illustrates the patchy space that
ensues if any constraint in addition to contiguity is used to define a
feasible solution.}
\label{fig:amoeba}
\end{figure}

To be sure, any graph theoretic framework compels a specific notion of
solution adjacency and proximity.  In particular, the number of edges
that must be changed to convert one solution to another defines the
distance between those two solutions in the state space.  In our graph
representation, the distance from one contiguous solution to another
contiguous solution is 1; and the set of all contiguous solutions
forms a connected set.  When feasibility includes another constraint,
this additionally constrained set of feasible solutions, using this
graph traversal mechanism, almost surely now comprises a disconnected
set.

\begin{figure}
\centering
\includegraphics[width=2.5in]{ex25.pdf}
\hspace{2.8cm}
\includegraphics[width=2.5in]{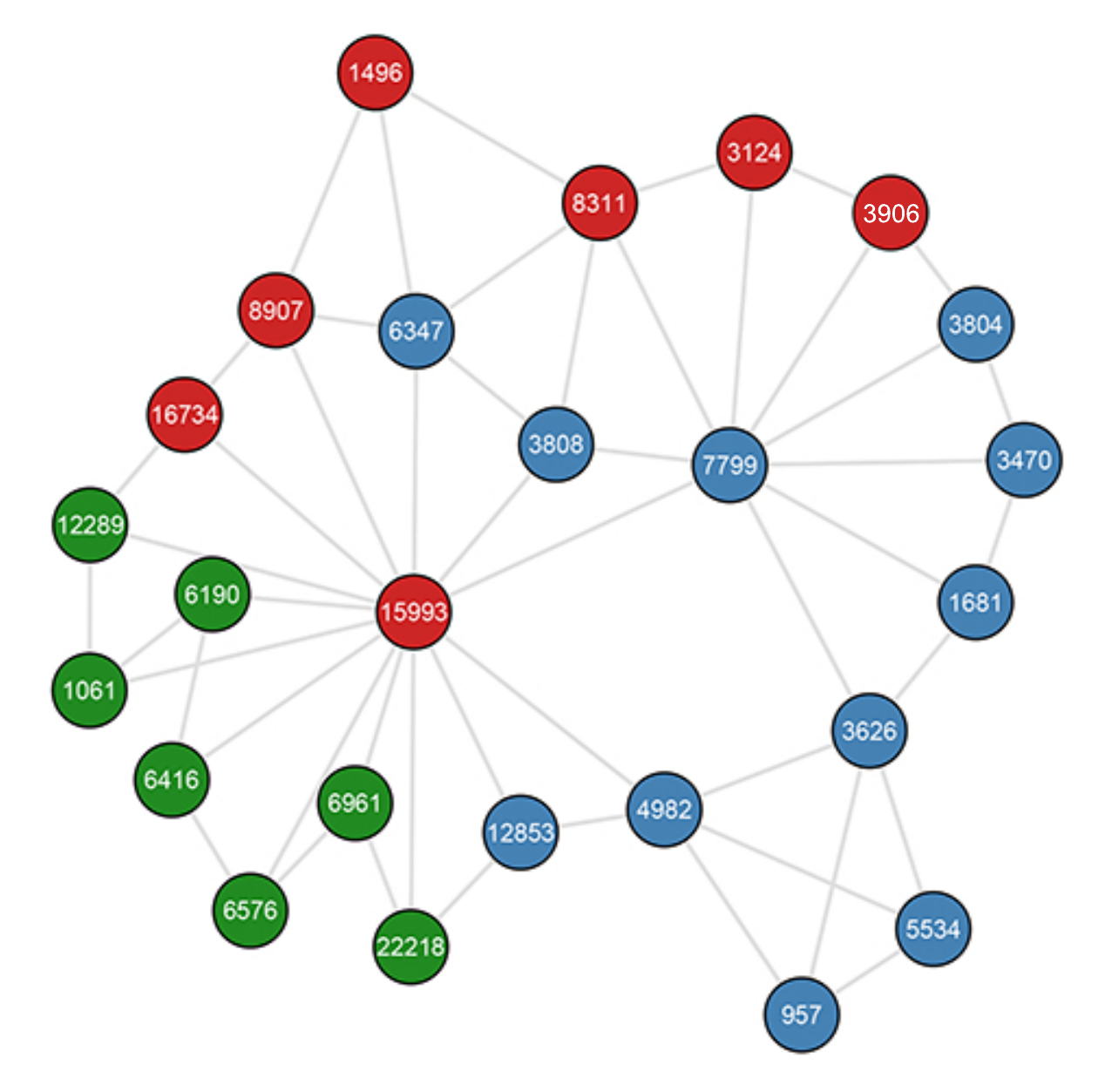}
\caption{Spatially Isolated Partitions.  With a Markov transition that
  involves only movement on zone boundaries, these two feasible
  partitions are isolated from all other feasible partitions.}
\label{fig:isolated}
\end{figure}

A connected space can be visualized with the amoeba shape shown in
Figure~\ref{fig:amoeba}.  The main feature of the ``shape'' of the
contiguous solutions is that it comprises a connected subspace such
that any contiguous partition can be reached from any other contiguous
partition with transitions that traverse only other contiguous
partitions.  That is, movement is completely confined within the
contiguous set, and movement into infeasible space is not necessary.
However, the imposition of any other constraint, say, weight
balancing, induces a ``patchy'' decision space that is not fully
connected by similar traversal.  We can visualize this as the figure
on the right in Figure~\ref{fig:amoeba}.  How to traverse from one
contiguous solution that satisfies the weight balancing criterion to
another contiguous solution that also satisfies the weight balancing
criteria is non-trivially more difficult, featuring not only greater
sparsity, but disconnected solution regions.  For our small example,
Figure~\ref{fig:isolated} shows two feasible partitions that are
proximate and reachable from the other, but isolated from {\em all}
other feasible partitions if traversal involves only unit swaps on
subgraph borders.

\subsection{Search Strategies}

As we have already noted, for the task of sampling solutions
satisfying only the contiguity constraint, a Markov transition like
ECMUT-1, as well as variants that have been proposed by
others~\citep{Bangiaetal:17}, is sufficient.  While the Markov chain
produced by this transition proposal is irreducible on the fully
connected state space, it is not irreducible when additional
constraints restrict the feasible solution space and induce a
disconnected state space.  A reducible Markov chain will not produce a
sample of the underlying space.

This issue can be resolved by 1) allowing movement into infeasible
space in the search for other feasible regions, 2) formulating a set
of parallel or serial chains that uniformly sample among and in the
disconnected regions, or 3) specifying a different transition proposal
that does produce an irreducible Markov chain.  The first option is
theoretically viable, but plainly produces massive computational
waste.  In our small example, the wasted effort is not strictly
limiting given the small problem size.  However, for a larger and more
realistic application, this approach is not practical without
enlisting greater computational capacity.  The second option avoids
the wasted computational effort spent wading through infeasible
regions, but then introduces a new problem to solve, that of
identifying a random sample of initial states.  For our small spatial
partitioning application, and for applications generally where the
state space is not continuous, parallel Markov chains can alleviate
some of the potential issues if we are able to generate random initial
states in different disconnected regions.  For our small empirical
application, generating these initial random states was relatively
easy, but for larger and more complex applications, devising a method
for identifying random initial states for a set of parallel chains is
non-trivial.  The third option requires ingenuity in devising the
transition proposal for the Markov chain.  The transition proposal
must be able to make large movements that transcend the
discontinuities that vex simpler and smaller transition
proposals.\footnote{To be sure, there may be other issues such as a
  multimodal state space replete with copious low and high energy
  states.  Here, there is a significant literature with many and
  varied proposals.  None of these can be regarded as a general fix.
  Rather, they must be tuned to the particular application.  The
  strategies generally fall into one of two camps.  The first is the
  set of methods that incorporate temperature as an auxiliary variable
  to facilitate sampling at a broad array of temperatures, both low
  and high.  Included in this set is, for example, parallel
  tempering~\citep{Geyer:91}, the equi-energy
  sampler~\citep{Kouetal:06}, and the Swendsen-Wang
  algorithm~\citep{SwendsenWang:87}.  The second is the set that
  incorporates information from past samples, which includes, for
  example, multicanonical sampling~\citep{BergNeuhaus:91}, the
  Wang-Landau algorithm~\citep{WangLandau:01} and stochastic
  approximation Monte Carlo~\citep{Liangetal:07}.  The choice depends
  on which strategy is most tightly coupled with the peculiarities of
  the specific state space.  This literature addresses issues of
  efficiency in the Markov chain.  That is, if one has an irreducible
  Markov chain that converges slowly, these strategies help to
  increase the efficiency of that chain and hasten convergence by
  producing a method that employs a mechanism for facilitating
  movement out of low energy states.  \citet{Fifieldetal:19} found
  that a parallel tempering approach was able to sample spatial
  partitions, but they failed to realize that the sample was enabled
  by the different chains that were initialized in different
  disconnected regions.  That is, what makes this approach tenable is
  not parallel tempering per se, but the ability to initiate different
  Markov chains (whether in parallel or not) at different and random
  starting states.}

Certainly the advantages of employing these separate options can made
more effective by combining the various strategies.  Movement need not
completely avoid infeasible solutions; parallelism can be implemented
from a variety of different serial scenarios; and larger Markov
transitions can be integrated along with parallel chains that
sometimes explore infeasible space.  Indeed, a confluence of these
strategies is likely to be more effective than any one on its own.
This is the approach that we employ in our EMCMC algorithm, where all
of these strategies have been incorporated.
 
Finally, computational scalability transcends the choice of search
strategy.  While convergence rates are certainly affected by the
efficiency of the transition proposal, for large applications, one
must also pursue computational strategies to further improve
efficiency.  This brings us to our last challenge, where even with a
significantly more efficient and irreducible Markov chain, the issue
of {\em scalability} remains.

\subsection{Computational Scalability with Massively Parallel Architecture}

\begin{figure}
\centering
\includegraphics[width=3in]{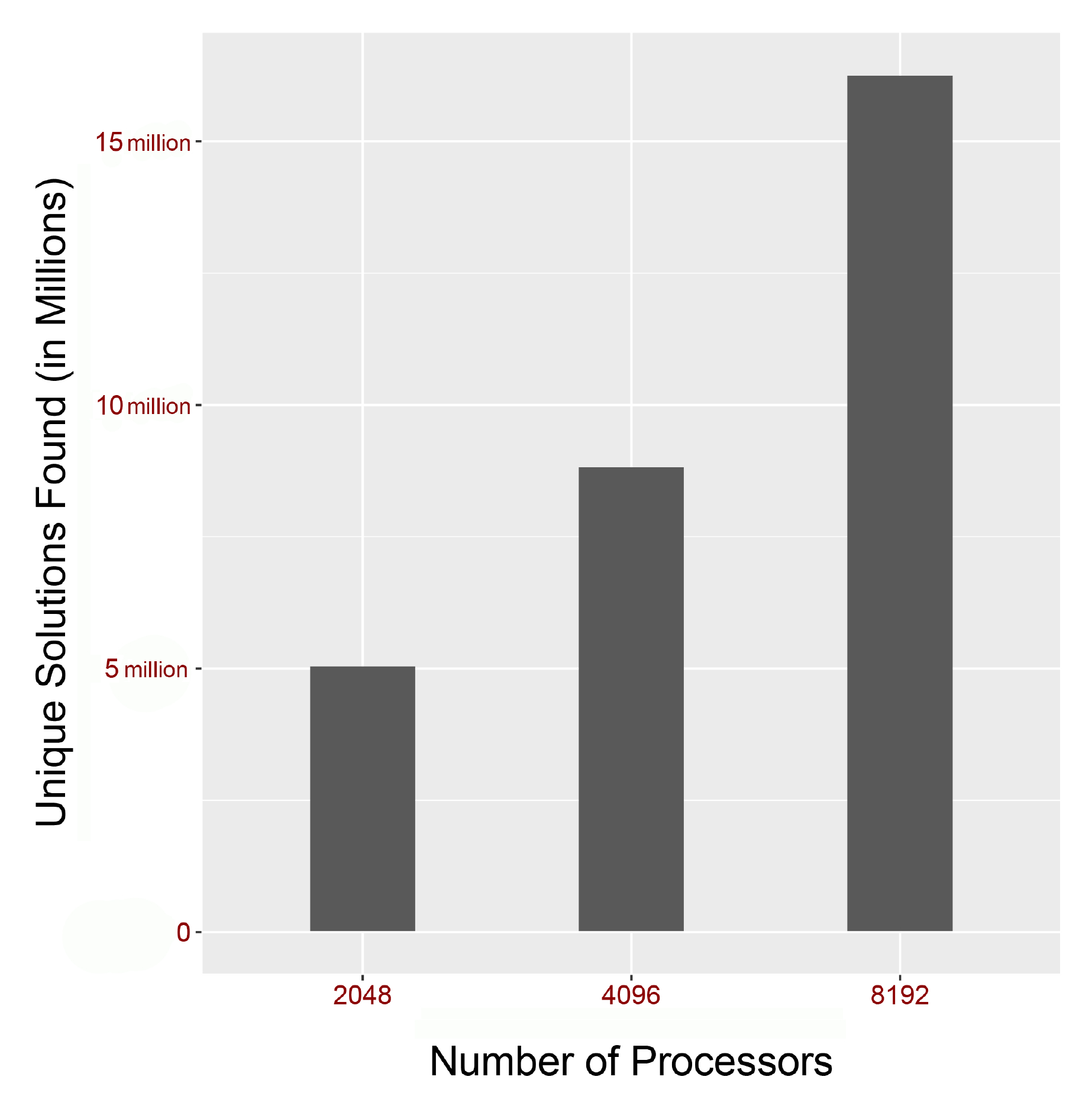}
\caption{Scalability of EMCMC Algorithm.  EMCMC is able to effectively
  scale with additional computational resources.}
\label{fig:scale}
\end{figure}

The nature of the EMCMC algorithm is already parallelized, with a
parallel structure that is straightforward to scale, so scalability is
already an integral part of our algorithm.  In Figure~\ref{fig:scale},
we see how our EMCMC algorithm scales when we enlist more processor
cores.  As we double the number of processor cores, EMCMC reaches an
increasingly larger portion of the underlying state space.  The figure
on the right shows the number of unique solutions identified in 10
minutes of computing time on the Blue Waters supercomputer.  When
8,192 processors are utilized, we explore almost 17 million unique
states, which is more than 3 times as many states as were explored
with 2,048 processor cores.  Importantly, with any number of processor
cores, the EMCMC algorithm was able to parse the underlying state
space to successfully produce a representative sample of the set of 
feasible solutions.

As the size of the state space grows larger, all of the challenges we
have identified must be addressed simultaneously and in increasingly
adept ways.  Indeed, as the application size grows, the effectiveness
and efficiency afforded by the spatial path relinking crossover
transition becomes increasingly essential.  The crossover transition
provides both larger movement, leading to greater efficiency and
effectiveness, as well as the ability to move between disconnected
portions of the overall state space.  Our directed crossover operator
guides intelligent and efficient space traversal while parallel chains
aid in diversity, reaching the different modes, and harnessing more
computing power.  In this small example, the power of PRCRX was not
evident because the space is sufficiently small that it can be
effectively sampled with the ECMUT operator and parallel chains that
were initiated at different random start states.  This strategy would
not be effective with larger applications where our spatial path
relinking crossover and our ability to harness massive computing power
becomes essential.

\section{Discussion}

We have devised a massively parallel Evolutionary Markov Chain Monte
Carlo (EMCMC) algorithm for sampling from astronomically large and
complex spatial state spaces.  The applicability and performance of
our algorithm has been demonstrated for a spatial partitioning
problem.  We have additionally scaled our algorithm to harness massive
computational power on parallel computing architecture, which is
necessary for problems of large scale.  The performance gain of our
algorithm is significant.

Within our EMCMC algorithm, we have introduced a transition rule for a
Markov chain that utilizes an optimization heuristic to enable large
movements.  Our optimization heuristic identifies local optimality
information via a directed search.  We use this information to
adapt and update the Markov chain in promising directions.  These
ideas stem from~\citet{Mezei:80} and~\citet{GelmanRubin:92}, who
suggested that efficiency in a Markov chain can be improved by
initially exploring where modes occur and then adjusting the proposal
function accordingly, as well as the work of~\citet{Liuetal:00} who
proposed the MTM method where any anchor point that is independent of
the current state can be used effectively in an MCMC sampler to direct
future draws.  In addition, our work builds upon our previous research
in the optimization realm, which we utilize to form the basis for an
adaptive deterministic procedure in our MCMC.

There remain a number of avenues for future exploration.  For
instance, the efficiency of the EMCMC sampler relies on a calibration
between the proposal step size, the size of the proposal set, $m$, and
the complexity of the landscape of the underlying distribution, $\pi$.
As well, while our EA operators were effective, there are other and
perhaps more efficient methods for landscape traversal in these
spatial state spaces.  Improvement in the optimization components
would have obvious translation to gains in performance for the
sampler.  There is an unknown and untested relationship between the
population size or the number of parallel chains and the rate of
convergence of the sampler.  A larger number of chains increases the
diversification of the search, which is desirable, while fewer but
longer chains have an impact on the convergence rate.  There are a
number of trade offs, some of which are application specific, while
others have roots in the theoretical properties of the underlying
model.

In addition, computational advances will improve the performance of
the sampler.  This includes innovations that increase the efficiency
of the computational algorithms as well as novel ways to adapt the
serial implementation of the algorithm to run on the massively
parallel architecture that is available on supercomputers.  This type
of architecture will have an obvious relationship with the population
size or the number of Markov chains that can be run in parallel.  It
has a more subtle relationship with how the search can be made more
efficient, by perhaps, intelligently increasing the diversity and the
intensity of the optimization heuristic in response to the variations
in state space landscape~\citep{ChoLiu:19}.  The large number of
processors could then span across the solution space to increase
diversity in the search while working collectively or intensively in
regions where the search is particularly difficult.

\section{Acknowledgements}
This research is part of the Blue Waters sustained petascale computing 
project, which is supported by the National Science Foundation 
(awards OCI-0725070 and ACI-1238993) and the State of Illinois. Blue Waters 
is a joint effort of the University of Illinois at Urbana-Champaign and 
its National Center for Supercomputing Applications.

\clearpage
\newpage
\bibliographystyle{apsr}
\bibliography{emcmc}

\end{document}